# SurgCBVR: A Multi-Domain Benchmark for Content-Based Surgical Video-to-Video Retrieval

Mohammad Eslami[1,2] Mousa Moradi[1] Negin Ghamsarian[3] Nazlee Zebardast[1]
Jennifer L Lindsey[1] Sophia Y. Wang[2] Lucia Sobrin[1] Rishi P Singh[1] Tobias Elze[1]
[1]Harvard Medical School [2]Stanford University [2]University of Bern
moha_esla@meei.harvard.edu

## Abstract

*Hospitals record thousands of hours of surgery every year, yet these archives remain almost impossible to search by content. A surgeon preparing a difficult case, reviewing a complication or teaching a resident cannot ask the archive "show me other operations or incidents that looked like this." Surgical computer vision has instead concentrated on frame-level classification (phase recognition, tool detection, segmentation), whose outputs are labels, not neighbors. We introduce SurgCBVR, the first benchmark for content-based video-to-video retrieval in surgery. It encompasses eight public datasets across four surgical domains—ophthalmic (cataract and MIGS), laparoscopic (cholecystectomy), robotic-assisted (prostatectomy and porcine-model training tasks), and colonoscopy—covering six procedure types and comprising 76,489 clips from 679 cases, totaling approximately 106 hours of video. SurgCBVR is organized as two tiers of increasing difficulty. Scene is retrieval of the same operative state (a surgical phase, a training task or a polyp's histology) across patients and sites, at three temporal granularities. Workflow is retrieval and localization of phase transitions and workflow deviations. Its 40 scene classes cover surgical phases, robotic training tasks and polyp histology, from 6–11 phases per procedure to lesion classes. The 74,765 Scene clips are sampled at three temporal granularities and three playback speeds. Also it contains 1,724 Workflow clips capture distinct phase-transition types. All clips share a fixed length and frame count, and every tier is evaluated against a phase-classifier oracle and a duration-only baseline, so that label and length shortcuts are measured rather than rewarded. We benchmark six generic and surgical foundation models, two video retrieval methods and contrastive heads trained on SurgCBVR, with retrieval restricted to the same procedure type and evaluated within datasets, across datasets and in pooled galleries. Frozen features barely exceed a timing-only baseline, transitions remain close to chance, and even trained heads lose up to half their accuracy across datasets, leaving ample room for surgery-aware video retrieval.*

## 1. Introduction

Minimally invasive and microscope-guided surgery have made the operating room a major source of medical video. Laparoscopic, robotic, endoscopic, and ophthalmic procedures routinely rely on live video for visualization, and substantial amounts of this footage are archived. These recordings capture how surgeons approach anatomical variation, perform complex maneuvers, and manage intraoperative challenges. Yet, despite their potential educational and clinical value, surgical video archives remain difficult to search. When searchable at all, retrieval typically depends on manually entered metadata or free-text descriptions rather than the visual content of the recordings.

**Importance and applications.** Surgical expertise is built through experience with cases. Residents learn by observing how experienced surgeons handle difficult dissections; departments review complications in morbidity-and-mortality conferences; video-based assessment compares surgical technique against reference performances; and surgeons preparing for challenging procedures may benefit from reviewing cases with similar anatomy or operative conditions. These applications share a fundamental requirement: given a surgical video clip, retrieve other clips depicting clinically similar situations. A phase classifier can identify a clip as showing phacoemulsification, but it cannot directly retrieve the most similar phacoemulsification segments from a large surgical archive. The potential value of such retrieval is particularly evident for rare but clinically significant events. For example, posterior capsule rupture is an important intraoperative complication of cataract surgery. A recent UK multicentre study of 961,208 cataract operations reported a posterior capsule rupture rate of 1.01% [28]. Management may require an-terior vitrectomy, modification of the intraocular lens implantation strategy, or other interventions depending on the circumstances of the rupture. Retrieving comparable events could help surgeons review alternative management strategies and allow trainees to study the operative conditions preceding complications. Such clinically meaningful retrieval is a long-term objective. Achieving it requires video representations that capture operative states and temporal workflow, together with standardized benchmarks for evaluating retrieval performance.

**The data gap.** Despite growing interest in surgical video understanding, content-based video-to-video retrieval remains relatively underexplored. Prior work on surgical video-to-video retrieval, notably Yu et al. [38], has demonstrated the feasibility of retrieving surgical content across procedures and events. However, a publicly reproducible, multi-domain benchmark with standardized relevance definitions, case-level splits, and cross-dataset evaluation remains lacking. More recent surgical video–language resources [4, 25, 40] have emphasized text-to-video retrieval, which addresses a different task from identifying visually similar intraoperative segments. Furthermore, clinically important events such as complications are infrequently annotated in publicly available datasets, partly because of their rarity and the practical and medico-legal challenges of collecting and sharing such data [16]. Existing annotations are therefore insufficient to support a large-scale, reproducible benchmark for incident retrieval (Sec. 3.7). SurgCBVR addresses the first gap by building a public, reproducible retrieval benchmark from openly licensed surgical datasets.

**Our approach.** Rather than restricting retrieval research to sparsely annotated complications, we build on the surgical annotations that are already widely available. Public surgical datasets provide temporal labels for phases, steps, actions, and other clinically meaningful events across multiple procedures and imaging modalities. We introduce SurgCBVR, a standardized benchmark that transforms these annotations into reproducible content-based video-to-video retrieval tasks with case-level splits and graded relevance.

Our primary focus is retrieval *within* a procedure type. Because procedure identity is generally available from clinical metadata, distinguishing cataract surgery from cholecystectomy is not the central challenge. Instead, the objective is to identify relevant operative moments among recordings of the same procedure, potentially acquired under different conditions. Cross-procedure retrieval is therefore outside the scope of this work. We also distinguish retrieval of annotated operative states from retrieval of clinically significant incidents (Sec. 3.7).

**Retrieval is not classification in disguise.** An important methodological concern is whether retrieval based on phase annotations simply reproduces a classification task. If relevance is determined solely by phase identity, a sufficiently accurate classifier could rank gallery clips by their predicted labels without learning a representation optimized for retrieval. We explicitly investigate this possibility by including a *phase-classifier oracle* baseline.

**Contributions.** We introduce SurgCBVR, the first publicly reproducible, multi-domain benchmark for content-based surgical video-to-video retrieval, comprising eight public datasets across four surgical domains and six procedure types, with 76,489 clips from 679 cases totaling approximately 106 hours of video. We formulate two complementary retrieval tiers: SCENE, which retrieves matching operative states across cases at three temporal granularities, and WORKFLOW, which retrieves and localizes phase transitions and workflow deviations. The benchmark includes 40 scene classes spanning surgical phases, robotic training tasks, and polyp histology, with 74,765 Scene clips sampled at three temporal granularities and three playback speeds, alongside 1,724 Workflow clips. We provide clips, standardized relevance annotations and case-level splits, and evaluate generic video encoders, surgical foundation models, phase-classification baselines, and trained retrieval methods to assess retrieval performance beyond label- and duration-based shortcuts.

## 2. Related Work

**Content-based video retrieval.** General-domain CBVR distinguishes duplicate, complementary and incident scene retrieval [13], with methods ranging from global clip descriptors to fine-grained spatio-temporal similarity learning [14] and distillation for efficient indexing [15]. Related formulations include partially relevant [6] and composed video retrieval [32]. These benchmarks are built from web video, where near-duplicates are common and visual diversity is large. Surgical video inverts both properties: duplicates never occur across patients, and the visual domain is narrow, so clinically distinct scenes can look alike while equivalent scenes look different. SurgCBVR adopts the query–gallery protocol of FIVR [13] but redefines relevance around clinical semantics.

**Surgical video understanding.** Public surgical datasets have driven phase recognition [9, 10, 17, 31, 37], action and triplet recognition [22, 26], gesture and skill analysis [8], and polyp detection in colonoscopy [5, 21]. Their evaluation is almost exclusively classification, detection or segmentation. SurgCBVR re-uses their annotations but asks a different question: whether learned representations place clinically similar content close together across patients and sites.

**Surgical retrieval and video–language models.** Yu *et al.* [38] introduced live laparoscopic video-to-video retrieval using uncertainty-aware compressed hashing, evaluated on cholecystectomy phases, gastric bypass phases, and surgical events across six procedure types. Although their work established an early benchmark for surgical retrieval, the underlying data were not publicly released, limiting reproducibility. We extend this direction through a publicly reproducible benchmark spanning eight datasets across four surgical domains (ophthalmic, laparoscopic, robotic-assisted, and colonoscopy), with broader task design encompassing graded relevance, multiple temporal granularities, cross-dataset evaluation, phase transitions, workflow deviations, and temporal localization. In parallel, surgical video–language pre-training approaches [4, 25, 40] have explored text-to-video retrieval, primarily using narrated surgical content. While these models provide relevant baselines, their ability to retrieve visually and temporally similar intraoperative segments remains insufficiently evaluated.

**Adverse events in surgical video.** Intraoperative adverse events are rarely annotated. Existing resources cover bleeding in laparoscopy [20, 24], simulated hemorrhage control [16], lens irregularities in cataract surgery [9], multi-type adverse events re-annotated on Cholec80 and robotic surgery [7], and graded events in gastric bypass [18]. Each is small, task-specific and annotated under its own schema; Sec. 3.7 discusses why, taken together, they do not yet support a retrieval benchmark.

# 3. The SurgCBVR Benchmark

## 3.1. Problem formulation

Let $\mathcal{V}$ be a collection of surgical videos, each belonging to a case $c(v)$, a procedure $p(v)$, a domain $d(v)$, a source dataset $D(v)$ and a site $s(v)$. A *clip* $x = (v, t_s, t_e)$ is a temporal window of $v$, rendered to a fixed length of 5 s (125 frames); windows longer than 5 s are played faster. Given a query clip $q$ and a gallery $\mathcal{G}$, a retrieval model $f$ maps clips to embeddings and ranks $\mathcal{G}$ by similarity $\mathrm{sim}(f(q), f(g))$. Each task defines a graded relevance function

$$\mathrm{rel}(q, g) \in \{0, 1, \dots, R\}, \tag{1}$$

where 0 denotes irrelevant. Two rules hold for every task: (i) gallery clips from the same case as the query, $c(g) = c(q)$, are removed from the ranking (neither positive nor negative), so that patient identity cannot be exploited; (ii) cases are split 40/20/40 into train/val/test within each source dataset, no case appears in more than one split, and queries and galleries are both drawn from the test split.

**Within-procedure galleries.** Every query is searched only against clips of the same procedure, $\mathcal{G}(q) = \{\, g : p(g) = p(q) \,\}$. This mirrors clinical use, where the procedure type is known from the operative record, and it removes trivially easy negatives that would inflate scores (a cataract query will never confuse a colonoscopy frame). Query and gallery clips are of the same kind: a SCENE query is searched among clips of its granularity, a speed-variant query among short (5 s) clips (Sec. 3.4), and a WORKFLOW query among transition clips.

**Evaluation settings.** *Within-dataset*: query and gallery come from the same source dataset. *Cross-dataset*: for procedures covered by two datasets, query and gallery come from different datasets, $D(g) \neq D(q)$, restricted to labels both datasets share; this measures robustness to a change of hospital, equipment and annotation team. *Pooled*: the gallery holds all clips of the procedure. The site $s(v)$ is recorded for future cross-site evaluation.

## 3.2. Domains and source datasets

SurgCBVR covers four domains with complementary imaging modalities: operating microscope (ophthalmic: cataract and glaucoma MIGS surgery), rigid endoscope (laparoscopic), robotic stereo endoscope (robotic) and flexible endoscope (colonoscopy). Two procedures are covered by two independent datasets each, phacoemulsification (Cataract-LMM, Cataract-1K) and cholecystectomy (Cholec80, HeiChole), which enables the cross-dataset setting. Table 1 summarizes the sources. Shared clip filenames encode the source dataset, video, tier, label, granularity and source start time, and `clips.csv` gives each clip's exact source window, case, split and playback speed. Every clip can thus be traced back to its source video, so that further annotations released with the source datasets (*e.g.* tool presence, anatomy or tissue labels) can be synchronized with SurgCBVR clips and used to define new relevance criteria. Retrieval methods may use only the pixel content; filenames and metadata are used for evaluation only.

## 3.3. Unified Annotation Layers

Source datasets differ in annotation granularity and terminology. To standardize their representation, we adopt a four-level hierarchy:

$$\underbrace{\text{domain}}_{d} \supset \underbrace{\text{procedure}}_{p} \supset \underbrace{\text{semantic class}}_{c} \supset \underbrace{\text{action}}_{a}. \tag{2}$$

Annotations from each dataset are converted into a unified, video-level table of temporal segments $(c, t_s, t_e)$, accompanied by case identifiers and site information when available. Idle periods, generic catch-all categories, and labels unsuitable for reliable cross-case retrieval (*e.g.* insufficient frequency or limited comparability) are excluded, with dataset-specific criteria detailed in the supplementary material.

The semantic-class level accommodates different annotation types: surgical phases or steps for most procedures, robotic training tasks for SurgVU's porcine-model sessions, and polyp histology (adenoma or hyperplastic) for REAL-Colon, which lacks conventional surgical phases. For colonoscopy, segments correspond to intervals during which an annotated lesion is visible; frames with insufficient lesion visibility, including lesions that are excessively small, enlarged, or truncated by image boundaries, are excluded. The action level and a separate *event* annotation layer, designed to capture event type, onset, offset, severity, and management, are reserved for future extensions and remain unpopulated in the current benchmark (Sec. 3.7). Datasets covering the same procedure are aligned to a common semantic-class vocabulary. Cross-dataset evaluation is restricted to shared classes, retaining 10 of 11 classes for phacoemulsification and all six for cholecystectomy.

Table 1. Source datasets in SurgCBVR. ✓ marks the tiers each dataset contributes to.

| Domain | Dataset | Procedure | SCENE | WORKFLOW |
|---|---|---|---|---|
| Ophthalmic | Cataract-1K [9] | phacoemulsification | ✓ | ✓ |
| | Cataract-LMM [1] | phacoemulsification | ✓ | ✓ |
| | Glaucoma MIGS [34] | 120° Goniotomy | ✓ | ✓ |
| Laparoscopic | Cholec80 [31] | Cholecystectomy | ✓ | |
| | HeiChole [33] | Cholecystectomy | ✓ | |
| Robotic-assisted surgery | GraSP [3] | Prostatectomies | ✓ | |
| | SurgVU [41] | Training tasks (porcine model) | ✓ | |
| Colonoscopy | REAL-Colon [5] | Colonoscopy | ✓ | |

Table 2. Temporal granularities of SCENE clips. Every clip is rendered to the same output length (5 s at 25 fps, *i.e.* 125 frames); granularities differ only in the source window $W$ compressed into it. $\tilde{D}$ is the median phase duration of the dataset (training split), $D$ the duration of the phase segment.

| Granularity | Level | Source window $W$ | Output |
|---|---|---|---|
| short | action | 5 s | 5 s, 1× |
| medium | step | $0.2\,\tilde{D} \in [7.5, 60]$ s | 5 s, compressed |
| phase | phase | $D/5$ | 5 s, compressed |
| speed-$s$ | robustness | $5s$ s, $s \in \{2, 3, 5\}$ | 5 s, $s$× |

### 3.4. Tier 1: SCENE retrieval

**Goal.** Retrieve clips showing the same operative state from other patients, surgeons and, where possible, sites. This tier calibrates the benchmark: it is where label-based shortcuts are expected to work.

**Why clip length needs care.** Two naive choices for clip length both fail across procedures. Fixed absolute lengths do not transfer: a 5 s clip covers a meaningful part of a cataract phase, which often lasts under two minutes, but a tiny and often uninformative fraction of a laparoscopic or robotic phase lasting tens of minutes. Lengths proportional to phase duration do transfer, but then the clip length itself encodes the phase—short phases yield short clips—so a model, or a trivial baseline, can match clips by duration rather than content. We resolve both by separating the *source window* $W$, the amount of surgery a clip covers, from the *output length*, which is the same for every clip.

**Clip construction.** For every phase segment $[t_s, t_e]$ of duration $D = t_e - t_s$, we sample clips around the phase centre $t_c = t_s + D/2$ with symmetric offsets proportional to duration,

$$\mathcal{O} = \left\{-\tfrac{2D}{10}, -\tfrac{D}{10}, 0, \tfrac{D}{10}, \tfrac{2D}{10}\right\}, \qquad t_{\text{start}} = t_c + o - \tfrac{W}{2}, \tag{3}$$

clamped to $[t_s, t_e - W]$, so that clips land in the core execution region of the phase. Table 2 lists the source windows. The *short* window is fixed at 5 s and kept at real speed. The *medium* window is set per dataset to $0.2\,\tilde{D}$, capped at 60 s; if it falls below 7.5 s (1.5× the short window) medium is omitted for that dataset, so it never duplicates short. Because $\tilde{D}$ is estimated on the training split, the window adapts to each procedure without looking at test cases. The *phase* window is $D/5$, discarded when $D/5 < 5$ s. For all three, segments with $D < 1.5W$ are skipped and duplicate starts after clamping are removed. To probe temporal robustness we add playback-speed variants of the centre clip, read over $W = 5s$ seconds and played at $s$× ($s \in \{2, 3, 5\}$), skipping segments with $D < W$. Colonoscopy clips are centred on annotated lesion appearances.

Every clip—medium and phase included—is then rendered to 5 s at 25 fps. All clips therefore have exactly 125 frames, whatever the dataset, phase or granularity, so neither duration nor frame rate can reveal the source (Sec. 3.8 reports a duration-only baseline that uses the original segment duration). Clips are re-encoded with frame-accurate seeking and their frame counts verified; we note that stream-copy cutting, a common shortcut, snaps cut points to keyframes and silently shortens clips by several seconds.

**Positives and negatives.** For a SCENE query $q$, a gallery clip $g$ of the same procedure is a *positive* if it carries the same label, $\phi(g) = \phi(q)$ (phase, training task or lesion class), and a *negative* otherwise. Clips from the query's own case are removed (Sec. 3.1), so every positive comes from another patient or session; queries left without a positive are skipped. mAP, Recall@$k$ and Precision@10 use this binary relevance.

**Graded relevance.** Neighboring phases share instruments and anatomy, so confusing them is a milder error than confusing distant phases. We therefore also report nDCG@10 with graded gains

$$\mathrm{rel}_1(q,g) = \begin{cases} 2 & \phi(g) = \phi(q) \\ 1 & \phi(g) \in \mathcal{N}(\phi(q)) \\ 0 & \text{otherwise,} \end{cases} \tag{4}$$

where $\mathcal{N}(\phi)$ is the set of phases adjacent to $\phi$ in the canonical workflow $\pi^\star$ (Sec. 3.5). Where no canonical order exists (lesion classes in colonoscopy), the gains are binary.

**Protocol.** Query and gallery clips share a granularity, giving one result per granularity; speed-$s$ queries are searched against the short gallery, as transformed queries in the spirit of FIVR [13]. Results are reported in the within-dataset, cross-dataset and pooled settings (Sec. 3.1).

### 3.5. Tier 2: WORKFLOW retrieval

**Goal.** Retrieve moments where the course of the operation changes. Transitions mark changes in surgical intent, instrument use and anatomical interaction

**Scope.** In this version, WORKFLOW is built for the ophthalmic procedures (Cataract-LMM, Cataract-1K, MIGS). Their phases are short and numerous (7–11 per procedure, some recurring), so each case yields many transitions whose neighboring phases both fit in a short window. Laparoscopic phases last minutes and give only about six transitions per case, and the robotic sessions in SurgVU are sequences of independent training tasks rather than a surgical workflow.

**Canonical workflow.** For each data set, we estimate the canonical phase order $\pi^\star$ by sorting phases by the median, over training cases, of the normalized time $t/T$ of their first occurrence ($T$ is the duration of the video). We also record the duration distribution of each phase on the training split. $\pi^\star$ defines the neighbors $\mathcal{N}(\phi)$ used in Eq. (4).

**Transition clips.** For a case, excluded segments (*e.g.* idle) are removed and consecutive segments with the same label are merged, giving an ordered phase sequence $(\phi_1, \ldots, \phi_n)$. Each consecutive pair defines a transition $\phi_i \to \phi_{i+1}$ at $t^{(i)}$, the midpoint of the gap $g_i \geq 0$ between $t_e(\phi_i)$ and $t_s(\phi_{i+1})$, which usually covers the instrument exchange. We extract the window $[\,t_e(\phi_i) - h,\ t_s(\phi_{i+1}) + h\,]$, centred on $t^{(i)}$, which holds the end of $\phi_i$, the gap and the start of $\phi_{i+1}$, with

$$h = \min\bigl(\Delta,\ \ell^{\text{last}}(\phi_i),\ \ell^{\text{first}}(\phi_{i+1})\bigr), \qquad \Delta = 5\,\text{s}, \tag{5}$$

where $\ell^{\text{last}}(\phi_i)$ and $\ell^{\text{first}}(\phi_{i+1})$ are the lengths of the last uninterrupted part of $\phi_i$ and the first uninterrupted part of $\phi_{i+1}$. The window therefore never reaches a third phase, even when a phase is very short (*e.g.* a 3 s viscoelastic injection gives $h = 3\,\text{s}$). A transition is skipped if $h < 2.5\,\text{s}$, if the window exceeds 20 s, or if the gap fills more than 40% of it, so that each phase occupies at least a quarter of the clip. As for SCENE, every window is rendered to 5 s at 25 fps (125 frames), *i.e.* played at $1\times$–$4\times$, so clip length cannot reveal that a neighboring phase is short. All transitions, including skipped ones, are released with their timestamps. Transitions of every type are kept, including rare ones (*e.g.* out-of-order or repeated phases) and each clip is tagged in the released metadata as regular or atypical (rare type, or closing an unusually long phase), for future analysis.

**Positives and relevance.** For a transition query, a gallery transition clip is a *positive* if it has the same transition type $\phi_i \to \phi_{i+1}$; mAP, Recall@$k$ and Precision@10 use this binary relevance. For nDCG@10 we use graded gains

$$\mathrm{rel}_2(q,g) = \begin{cases} 2 & \text{same transition type} \\ 1 & \text{same phase before or same phase after} \\ 0 & \text{otherwise.} \end{cases} \tag{6}$$

We report results over all queries.

**Towards moment localization.** Retrieving pre-cut clips understates the real task, in which the archive consists of full recordings. Because every transition is released with its timestamp in the untrimmed video, SurgCBVR also supports a localization variant: given a transition query, return a temporal window in each untrimmed gallery video, counted as correct if its temporal IoU with a same-type ground-truth transition exceeds $\tau$. We leave its evaluation to future work.

### 3.6. Splits and Statistics

Extracted clips from each dataset are partitioned into training, validation, and test sets (40/20/40) at the case level, independently of any official dataset splits. All clips originating from the same case are assigned to a single split to prevent data leakage. Dataset-specific parameters used in benchmark construction, including the medium-window duration, canonical workflow order $\pi^\star$, and overlong-phase thresholds, are estimated exclusively from the training split. The training set supports retrieval model development, while the validation set is used for model selection and hyperparameter tuning. All reported performance results are obtained on the held-out test set. Case-level assignments are fixed and will be preserved in future benchmark versions, allowing newly introduced annotations to inherit the same splits. Table 3 summarizes the statistics for each retrieval tier (test set only).

Table 3. SurgCBVR statistics (test split). #Queries: test clips of the subset. #Gallery: test clips searched (summed over procedures; each query is ranked only against its own procedure, without clips of its own case). Speed-$s$ ($s = 2, 3, 5$) queries are searched in the short gallery. #Classes: phases, tasks or lesion classes (SCENE) and transition types (WORKFLOW), counted per procedure.

| Tier | Subset | #Queries | #Gallery | #Classes | #Datasets |
|---|---|---|---|---|---|
| SCENE | short | 11,435 | 11,435 | 40 | 8 |
| | medium | 3,775 | 3,775 | 20 | 4 |
| | phase | 7,290 | 7,290 | 38 | 7 |
| | speed-$s$ | 5,341 | 11,435 | 40 | 8 |
| WORKFLOW | transition | 751 | 751 | 65 | 3 |

### 3.7. Why incident retrieval is not included in this version

Retrieval of intraoperative adverse events, such as posterior capsule rupture (PCR) in cataract surgery, is the clinical target that motivates SurgCBVR. We do not build it from existing public data for three reasons. First, such events are rare (PCR occurs in about 2% of operations [11]), so public datasets contain too few independent cases for case-level splits to give stable scores. Second, the available incident annotations follow incompatible schemas (masks vs. clip labels, severity vs. presence, differing onset definitions) [7, 18, 20, 24, 27], so pooling them would mix non-equivalent events. Third, some announced incident datasets are not yet fully available for redistribution. A credible incident tier therefore requires new expert annotation under one schema, which we leave to a future version. SurgCBVR is prepared for it: the schema reserves an event layer (Sec. 3.3), splits are fixed per case so new labels inherit them, and per-procedure galleries let incident queries be added with hard negatives from uneventful cases without altering existing tasks.

### 3.8. Baselines

All methods score a query–gallery pair by the cosine similarity of $\ell_2$-normalised clip embeddings unless stated otherwise. Encoders are used frozen with their public weights.

**Clip embeddings from frozen encoders.** Every clip is 5 s long (125 frames at 25 fps). An encoder that takes single images is applied to frames sampled across the clip; an encoder that takes short videos is applied to consecutive windows of its native input length covering the clip. The frame or window features are then combined into one clip embedding by averaging them (*mean*) or by taking their element-wise maximum (*max*), and we report both. Specifically:

- DINOv2 ViT-B/14 [12, 23] (image): the CLS token of 25 frames (one every 0.2 s).
- C3D [30] (video, Sports-1M): fc7 features of 7 windows of 16 consecutive frames.

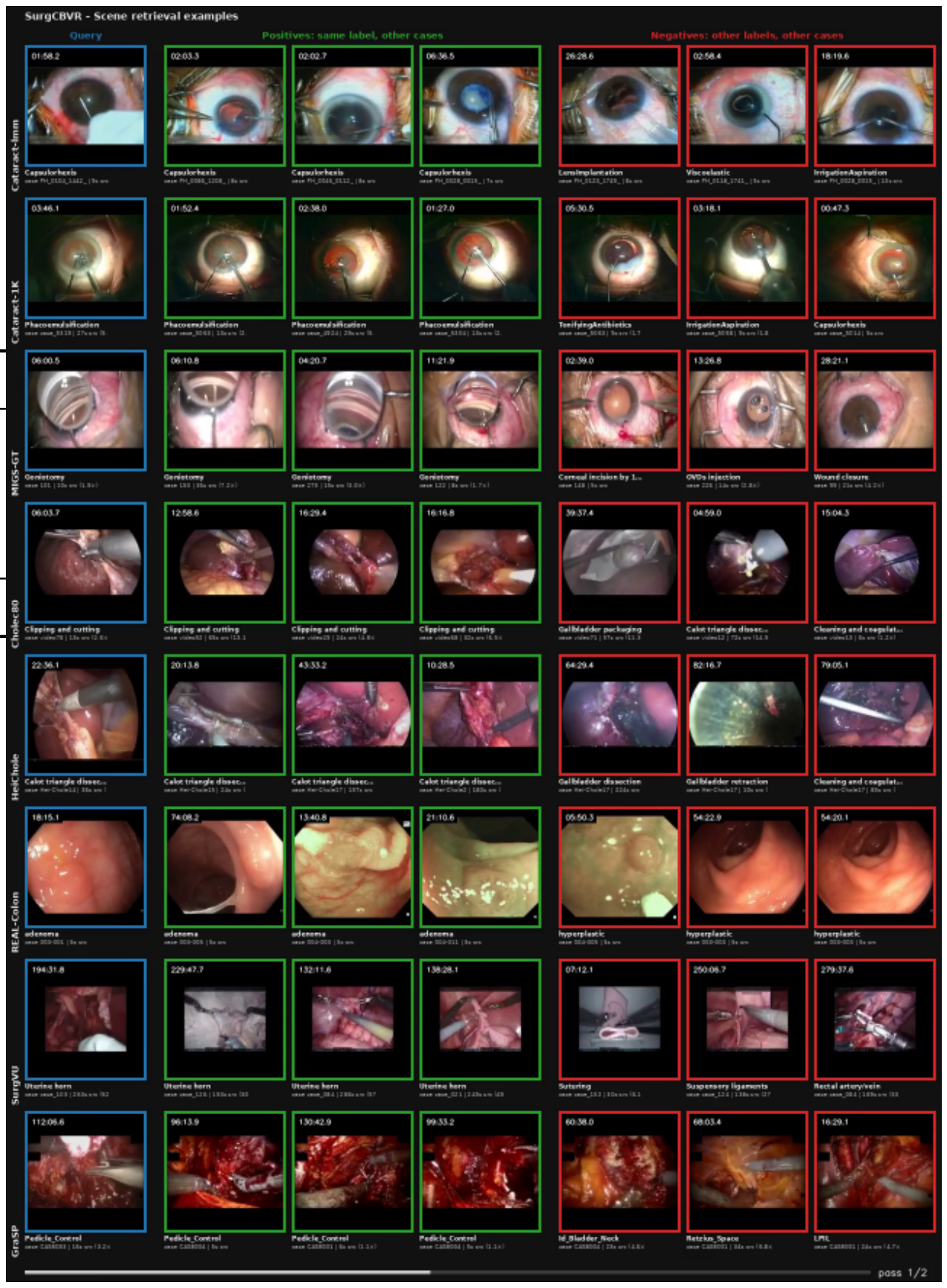

Figure 1. Representative frames from the SCENE retrieval task across datasets, illustrating query clips and their corresponding positive and negative samples. The associated video examples are provided in the supplementary material (`https://youtu.be/v4oZiq-o2I0`) (best view 300% magnification). More figures are available in supplementary material.

- VideoMAE V2 ViT-B [29, 35] (video, Kinetics-710): 3 windows of 16 frames (every third frame); mean or max over all tokens of all windows.
- V-JEPA 2 [2], InternVideo2 [36], SurgVLP/PeskaVLP [39, 40] and SurgRec-MAE [19].

The first three are generic image and video encoders; SurgVLP, PeskaVLP and SurgRec-MAE are surgical foundation models.

**Video retrieval methods.** (i) ViSiL [14], used as published with its pretrained weights: frames at 1 fps, region features, and its learned frame-to-frame similarity network. It was trained for near-duplicate and same-event retrieval, not for surgical phases. (ii) Frame matching (chamfer): a training-free similarity in the spirit of ViSiL on DINOv2 features of 16 frames per clip; each query frame is matched to its most similar gallery frame and the matches are averaged (we also report the symmetric version).

**Trained on SurgCBVR.** (i) Contrastive retrieval head

(SupCon): a two-layer MLP on frozen features of each encoder, trained on the training split of each dataset with a supervised contrastive loss, where clips with the same label are positives and pairs from the same case are ignored; model selection uses validation mAP. (ii) Phase-classifier oracle: a one-hidden-layer MLP (512 units) on the same frozen features, trained to predict the label on each dataset's training split. A query is ranked with the classifier of its own dataset, applied to the query and to all gallery clips, so in the cross-dataset setting the classifier has never seen the gallery dataset. We rank (a) by cosine similarity of its hidden-layer features and (b) by predicted label, the shortcut a retrieval benchmark should expose.

**Reference baselines.** Random ranking gives chance level. Duration only ranks the gallery by $-|D(q) - D(g)|$, the difference in the original duration of the source segment (transition window for WORKFLOW), which the rendered clips do not reveal; it measures how much could be obtained from timing alone.

## 4. Results

We report mAP (%) on the test split; queries and gallery clips both come from test cases (Sec. 3.1). Table 4 gives the main results, using mean pooling for all encoders (max pooling is similar or worse Table 13). Due to space limits, the remaining results are in the supplementary material: Recall@$k$ and nDCG@10, per-dataset results, the medium granularity, max pooling and confidence intervals.

**Frozen representations barely exceed timing.** Off-the-shelf encoders reach 31–41% average mAP on short clips, against 23% for chance and 30.5% for the duration-only reference, which ranks by the length of the source segment, a quantity no method can observe. On phacoemulsification, only one frozen encoder, VideoMAE V2 (22.1%), beats duration only (18.7%). Generic embeddings are dominated by appearance shared across a case or a setup (illumination, microscope, anatomy) rather than by the surgical phase, so similar-looking clips of different phases are ranked above same-phase clips from other patients. Surgical pretraining helps only within its own domain: PeskaVLP is the best frozen encoder (41.1%), with clear gains on cholecystectomy (45.9%) and prostatectomy (43.0%), but no gain on the ophthalmic procedures; the surgical video encoder SurgRec-MAE is below the generic DINOv2 when frozen.

**Video retrieval methods.** Pretrained ViSiL, never trained on surgery, is on par with the best generic encoders on SCENE (37.9%) and is the best frozen method on WORKFLOW (19.1% and 45.7% vs. at most 16.3% and 31.4%). Matching DINOv2 frames by chamfer similarity without learning, in contrast, changes little over mean pooling (35.6% vs. 34.9%). Fine-grained spatio-temporal matching thus helps where order matters (a transition is phase *A then B*), not where a single phase must be recognised.

**Learning on SurgCBVR closes much of the gap.** A small contrastive head on frozen features roughly doubles mAP (VideoMAE V2: 36.4→67.0%), so the relevant information is present in the features but not organised by surgical meaning. The ranking of encoders changes: the generic video encoders VideoMAE V2 and V-JEPA 2 gain most and overtake PeskaVLP, so frozen performance is a poor proxy for how useful a representation is. The benchmark is far from saturated except for SurgVU (91%), whose four training tasks differ strongly in setup.

**Ranking with a classifier.** We also rank clips with a phase classifier trained on the same features. Comparing clips by the classifier's predicted probabilities works as well as the contrastive head (VideoMAE V2: 66.8% vs. 67.0%), and this holds for every encoder (Table 20). Using only the predicted label is about ten points worse (56.3%), and using the classifier's hidden features is worse still (51.9%). What matters is training with the benchmark labels, and then comparing clips by their soft predictions, not their hard labels.

**Granularity and playback speed.** Longer temporal context helps every method. Averaged over the five procedures with phase-level clips (all but colonoscopy), phase clips score 4–10 points above short clips (*e.g.* VideoMAE V2 with the head: 70.9% short vs. 78.0% phase, Table 11), and medium clips lie in between. Playback-speed changes have relatively small effects on frozen encoders: over the same five procedures, frozen encoders are unchanged (VideoMAE V2: 33.4% at 1×, 34.2% at 5×) and SupCon heads improve slightly (70.9% → 75.3%), since a faster clip covers more of the phase. Within this range, retrieval is driven by content rather than motion speed.

**WORKFLOW is the hardest tier.** Frozen encoders achieve 11.7–16.3% mAP on phacoemulsification against 9.8% for chance, and 28.7–31.4% on goniotomy against 21.4% for chance. The best trained head reaches 40.7% and 53.9%, respectively (Table 16).

**Cross-dataset retrieval exposes a domain gap.** (Table 4) Searching the other dataset of the same procedure, with shared labels only, costs the best method 13 points on cholecystectomy (66.7→53.9%) but halves it on phacoemulsification (64.7→33.6%), where frozen encoders fall to near chance (11.3–14.3% vs. 9.8%) and below duration only. The phase classifiers show the same asymmetry: trained on Cataract-LMM, the VideoMAE V2 classifier reaches 72% balanced accuracy on Cataract-LMM but 30% on Cataract-1K (Cholec80→HeiChole: 84→64%). Although both cataract datasets use the same phase names, differences in recording setup and in where annotators place phase boundaries make transfer much harder than within-dataset results suggest, which is the robustness gap SurgCBVR is designed to measure.

**Limitations and outlook.** Lesion histology is the hard-est SCENE task. Distinguishing adenomas from hyperplastic polyps by appearance is difficult even for experienced endoscopists: it depends on subtle surface and vascular patterns, usually inspected at close range or with narrow-band imaging, whereas a 5 s clip from routine footage often shows the lesion only briefly or at a distance. Correspondingly (Table 10), no method is reliably above chance (45.5%), and trained heads do not improve on frozen features with only 50 training clips. In laparoscopy, the camera occasionally leaves the abdomen, most often during gallbladder extraction (labeled *gallbladder retraction*) and for lens cleaning; such moments remain in the current release. The prostatectomy test split contains three patients, so its results are indicative, and SurgVU, whose training tasks differ strongly in setup, is close to saturation once a head is trained. We report colonoscopy and prostatectomy as pilot procedures; larger releases, lesion-centred annotations and out-of-body filtering are natural next steps for these parts of the benchmark.

Table 4. Retrieval mAP (%) on the SurgCBVR test split. SCENE, short clips, within-dataset galleries, per procedure and averaged over all six (*Avg*); *Phase*: phase-level clips averaged over the five procedures that have them; WORKFLOW: transition clips, within dataset; *Cross-dataset*: short clips, gallery from the other dataset of the same procedure. Bold: best non-reference method per column. †Uses the duration of the source segment, which the clips do not reveal. ‡500 random queries per matrix.

| | SCENE (within dataset) | | | | | | | | WORKFLOW | | Cross-dataset | |
|---|---|---|---|---|---|---|---|---|---|---|---|---|
| Method | Phaco | Gonio. | Chole. | Prost. | SurgVU | Colon | Avg | Phase | Phaco | Gonio. | Phaco | Chole. |
| *Reference* | | | | | | | | | | | | |
| Random (chance) | 11.0 | 15.7 | 17.9 | 13.8 | 33.9 | 45.5 | 23.0 | 20.6 | 9.8 | 21.4 | 9.8 | 17.8 |
| Duration only† | 18.7 | 26.1 | 30.6 | 17.0 | 38.9 | 51.5 | 30.5 | 27.9 | 12.4 | 25.4 | 18.7 | 29.1 |
| *Frozen encoders* | | | | | | | | | | | | |
| C3D | 14.8 | 25.2 | 28.4 | 25.2 | 40.9 | **60.5** | 32.5 | 33.6 | 12.7 | 28.7 | 11.8 | 25.4 |
| DINOv2 | 16.2 | 32.7 | 31.4 | 31.9 | 48.3 | 48.9 | 34.9 | 38.4 | 12.6 | 30.2 | 12.2 | 27.8 |
| VideoMAE V2 | 22.1 | 35.4 | 30.4 | 31.2 | 48.1 | 51.4 | 36.4 | 40.7 | 16.3 | 31.4 | 14.3 | 26.4 |
| V-JEPA 2 | 15.7 | 29.0 | 32.1 | 36.0 | 44.7 | 54.9 | 35.4 | 39.6 | 13.9 | 29.4 | 12.0 | 29.5 |
| PeskaVLP | 17.5 | 33.1 | 45.9 | 43.0 | 52.4 | 54.9 | 41.1 | 44.7 | 14.8 | 29.0 | 12.6 | 40.0 |
| SurgRec-MAE | 13.9 | 25.6 | 25.6 | 23.5 | 43.5 | 54.4 | 31.1 | 31.3 | 11.7 | 30.1 | 11.3 | 24.3 |
| *Video retrieval* | | | | | | | | | | | | |
| Chamfer (DINOv2 frames) | 16.9 | 34.1 | 31.9 | 32.5 | 49.5 | 48.8 | 35.6 | 39.2 | 13.3 | 31.6 | 12.2 | 28.2 |
| ViSiL ‡ | 20.3 | 40.1 | 30.5 | 31.0 | 50.4 | 55.2 | 37.9 | 38.5 | 19.1 | 45.7 | 17.1 | 27.5 |
| *Trained on SurgCBVR (SupCon head)* | | | | | | | | | | | | |
| C3D | 26.2 | 47.2 | 42.2 | 44.4 | 76.1 | 46.2 | 47.0 | 56.9 | 19.0 | 31.4 | 12.4 | 32.1 |
| DINOv2 | 47.3 | 59.8 | 63.1 | 64.6 | 91.3 | 43.5 | 61.6 | 73.6 | 27.1 | 38.4 | 27.6 | 48.5 |
| VideoMAE V2 | **64.7** | **68.1** | **66.7** | 64.6 | 90.4 | 47.2 | **67.0** | 78.0 | **40.7** | **53.9** | 33.6 | **53.9** |
| V-JEPA 2 | 58.0 | 64.8 | 65.4 | **68.1** | **91.3** | 51.3 | 66.5 | **78.2** | 37.2 | 52.8 | **34.4** | 51.0 |
| PeskaVLP | 40.9 | 60.3 | 64.0 | 62.5 | 89.1 | 50.5 | 61.2 | 71.6 | 25.8 | 39.3 | 18.4 | 49.8 |
| SurgRec-MAE | 39.4 | 56.3 | 48.8 | 61.7 | 85.4 | 51.9 | 57.2 | 67.1 | 23.4 | 50.6 | 15.7 | 39.1 |
| *Classifier oracle (VideoMAE V2)* | | | | | | | | | | | | |
| hidden features | 39.5 | 56.5 | 48.1 | 47.7 | 70.4 | 49.2 | 51.9 | 60.7 | 23.2 | 32.0 | 27.2 | 39.4 |
| predicted label | 52.8 | 55.8 | 49.9 | 53.0 | 81.0 | 45.5 | 56.3 | 67.9 | 27.8 | 30.0 | 15.0 | 38.1 |

## 5. Limitations and Ethics

**Data and licenses.** SurgCBVR uses only public, de-identified datasets. Where licenses permit, we redistribute processed clips and annotations under their original licenses, with attribution and documented modifications (`github:xx;huggingface:xx`). Otherwise, we provide annotations, clip boundaries, and reproduction scripts. Given source licensing restrictions, SurgCBVR is intended for non-commercial research. **Scope.** SurgCBVR evaluates operative-state and workflow retrieval, not incident retrieval (Sec. 3.7). Cross-procedure retrieval is excluded because procedure types are documented clinically. Fixed-length clips couple temporal context and playback speed at longer granularities (Sec. 4). Results from small test splits are indicative. **Clinical use.** SurgCBVR assesses representation quality, not clinical safety; clinical deployment requires prospective validation.

## 6. Conclusion

We introduced SurgCBVR, a benchmark for content-based video-to-video retrieval in surgery that spans four domains, six procedure types and eight public datasets, with SCENE retrieval at several temporal granularities and playback speeds and WORKFLOW retrieval of phase transitions. Retrieval is restricted to the same procedure type and evaluated within datasets, across datasets and in pooled galleries.

Our baselines show that frozen generic and surgical foundation models barely exceed a timing-only reference, that transitions remain close to chance, and that a lightweight head trained on SurgCBVR roughly doubles accuracy but still loses up to half of it across datasets. Surgery-aware, order-sensitive and dataset-robust video representations are therefore far from solved. We release clips, annotations and code to support this work, and plan to extend SurgCBVR with expert-annotated intraoperative incidents.

**LLM Usage Statement.** Large language models (LLMs) were used for proofreading, grammar correction, minor script development, and code review to identify potential errors and inconsistencies. No LLMs contributed to research ideation, experimental design, or substantive manuscript writing. All code and experimental results were reviewed and verified by the authors.

# SurgCBVR: A Multi-Domain Benchmark for Content-Based Surgical Video-to-Video Retrieval

## Supplementary Material

Table 5. SCENE retrieval mAP (%), Phacoemulsification, within-dataset galleries, per granularity. Speed-$s$ queries are searched in the short gallery. Short-clip queries: 975 from 23 test cases. †Ranks by the duration of the source segment, which the clips do not reveal. ‡500 random queries per matrix. Bold: best non-reference method per column.

| Method | Short | Phase | 2× | 3× | 5× |
|---|---|---|---|---|---|
| *Reference* | | | | | |
| Random (chance) | 11.0 | 16.5 | 11.0 | 10.9 | 10.9 |
| Duration only† | 18.7 | 24.8 | 18.9 | 19.7 | 23.1 |
| *Frozen encoders (mean pooling)* | | | | | |
| C3D | 14.8 | 24.8 | 14.9 | 14.9 | 15.6 |
| DINOv2 | 16.2 | 23.9 | 16.2 | 16.0 | 16.6 |
| VideoMAE V2 | 22.1 | 31.1 | 22.6 | 21.9 | 21.5 |
| V-JEPA 2 | 15.7 | 25.4 | 15.6 | 15.4 | 15.6 |
| PeskaVLP | 17.5 | 24.6 | 17.3 | 16.7 | 17.1 |
| SurgRec-MAE | 13.9 | 22.1 | 13.8 | 13.6 | 13.8 |
| *Video retrieval* | | | | | |
| Chamfer (DINOv2) | 16.9 | 24.8 | 16.9 | 16.8 | 17.6 |
| Sym. chamfer (DINOv2) | 16.5 | 24.3 | 16.7 | 16.7 | 17.7 |
| ViSiL‡ | 20.3 | 29.0 | – | – | – |
| *SupCon head (trained)* | | | | | |
| C3D | 26.2 | 38.7 | 27.3 | 28.0 | 31.0 |
| DINOv2 | 47.3 | 56.9 | 48.2 | 48.9 | 52.5 |
| VideoMAE V2 | **64.7** | **70.0** | **66.5** | **66.9** | **69.6** |
| V-JEPA 2 | 58.0 | 64.9 | 59.7 | 60.1 | 62.7 |
| PeskaVLP | 40.9 | 49.5 | 41.7 | 41.8 | 45.4 |
| SurgRec-MAE | 39.4 | 50.8 | 40.8 | 40.6 | 44.8 |
| *Classifier oracle: hidden features* | | | | | |
| C3D | 18.6 | 29.5 | 18.9 | 18.6 | 19.8 |
| DINOv2 | 29.0 | 37.7 | 29.1 | 28.8 | 30.2 |
| VideoMAE V2 | 39.5 | 46.7 | 40.1 | 39.1 | 38.8 |
| V-JEPA 2 | 29.0 | 40.0 | 29.9 | 29.6 | 31.6 |
| PeskaVLP | 24.2 | 31.1 | 24.1 | 23.1 | 24.1 |
| SurgRec-MAE | 24.0 | 33.7 | 24.0 | 23.0 | 23.2 |
| *Classifier oracle: predicted label* | | | | | |
| C3D | 19.4 | 30.6 | 19.9 | 20.3 | 21.9 |
| DINOv2 | 37.5 | 48.0 | 38.0 | 38.3 | 41.2 |
| VideoMAE V2 | 52.8 | 59.1 | 54.6 | 54.4 | 55.9 |
| V-JEPA 2 | 44.3 | 50.6 | 44.8 | 44.3 | 45.7 |
| PeskaVLP | 28.1 | 34.0 | 27.7 | 26.8 | 29.3 |
| SurgRec-MAE | 29.1 | 41.0 | 29.4 | 29.1 | 30.9 |

Table 6. SCENE retrieval mAP (%), Goniotomy (MIGS), within-dataset galleries, per granularity. Speed-$s$ queries are searched in the short gallery. Short-clip queries: 2,035 from 59 test cases. †Ranks by the duration of the source segment, which the clips do not reveal. ‡500 random queries per matrix. Bold: best non-reference method per column.

| Method | Short | Phase | 2× | 3× | 5× |
|---|---|---|---|---|---|
| *Reference* | | | | | |
| Random (chance) | 15.7 | 21.5 | 16.0 | 16.5 | 17.1 |
| Duration only† | 26.1 | 29.0 | 26.2 | 26.5 | 27.7 |
| *Frozen encoders (mean pooling)* | | | | | |
| C3D | 25.2 | 38.8 | 26.1 | 28.2 | 30.8 |
| DINOv2 | 32.7 | 45.5 | 32.9 | 35.2 | 37.7 |
| VideoMAE V2 | 35.4 | 47.7 | 35.9 | 37.4 | 38.9 |
| V-JEPA 2 | 29.0 | 42.6 | 30.3 | 32.9 | 35.5 |
| PeskaVLP | 33.1 | 44.7 | 33.5 | 35.5 | 37.2 |
| SurgRec-MAE | 25.6 | 34.2 | 26.2 | 27.4 | 27.9 |
| *Video retrieval* | | | | | |
| Chamfer (DINOv2) | 34.1 | 46.9 | 34.4 | 36.9 | 39.7 |
| Sym. chamfer (DINOv2) | 32.6 | 43.7 | 32.7 | 34.8 | 37.0 |
| ViSiL‡ | 40.1 | 51.8 | – | – | – |
| *SupCon head (trained)* | | | | | |
| C3D | 47.2 | 61.6 | 48.8 | 52.6 | 57.5 |
| DINOv2 | 59.8 | 72.3 | 60.9 | 65.7 | 69.7 |
| VideoMAE V2 | **68.1** | 77.2 | **69.1** | **72.2** | **75.8** |
| V-JEPA 2 | 64.8 | **77.3** | 66.3 | 70.2 | 74.6 |
| PeskaVLP | 60.3 | 71.1 | 61.6 | 66.0 | 69.5 |
| SurgRec-MAE | 56.3 | 69.3 | 56.9 | 61.2 | 65.8 |
| *Classifier oracle: hidden features* | | | | | |
| C3D | 37.0 | 52.1 | 38.4 | 41.7 | 45.8 |
| DINOv2 | 48.9 | 61.4 | 49.7 | 53.1 | 56.4 |
| VideoMAE V2 | 56.5 | 69.8 | 58.1 | 61.3 | 64.1 |
| V-JEPA 2 | 48.3 | 63.7 | 50.4 | 54.4 | 58.4 |
| PeskaVLP | 44.1 | 55.5 | 45.0 | 47.7 | 50.3 |
| SurgRec-MAE | 43.9 | 57.1 | 45.8 | 48.6 | 50.5 |
| *Classifier oracle: predicted label* | | | | | |
| C3D | 39.5 | 54.1 | 40.6 | 44.2 | 48.1 |
| DINOv2 | 46.9 | 58.6 | 47.4 | 51.4 | 54.6 |
| VideoMAE V2 | 55.8 | 66.5 | 56.7 | 60.3 | 63.8 |
| V-JEPA 2 | 50.5 | 65.0 | 51.9 | 55.6 | 59.4 |
| PeskaVLP | 48.4 | 58.5 | 48.8 | 51.5 | 54.7 |
| SurgRec-MAE | 43.2 | 58.2 | 44.7 | 47.6 | 50.4 |

Table 7. SCENE retrieval mAP (%), Cholecystectomy, within-dataset galleries, per granularity. Speed-$s$ queries are searched in the short gallery. Short-clip queries: 895 from 30 test cases. †Ranks by the duration of the source segment, which the clips do not reveal. ‡500 random queries per matrix. Bold: best non-reference method per column.

| Method | Short | Medium | Phase | 2× | 3× | 5× |
|---|---|---|---|---|---|---|
| *Reference* | | | | | | |
| Random (chance) | 17.9 | 18.2 | 17.9 | 17.8 | 17.6 | 17.8 |
| Duration only† | 30.6 | 30.1 | 30.1 | 30.6 | 30.4 | 30.3 |
| *Frozen encoders (mean pooling)* | | | | | | |
| C3D | 28.4 | 30.0 | 34.1 | 28.1 | 28.1 | 28.2 |
| DINOv2 | 31.4 | 35.5 | 36.0 | 31.8 | 32.2 | 32.9 |
| VideoMAE V2 | 30.4 | 33.9 | 35.8 | 29.9 | 30.1 | 30.6 |
| V-JEPA 2 | 32.1 | 36.2 | 40.2 | 31.6 | 31.2 | 30.7 |
| PeskaVLP | 45.9 | 52.6 | 53.4 | 46.6 | 47.6 | 49.5 |
| SurgRec-MAE | 25.6 | 27.1 | 27.0 | 25.7 | 25.4 | 25.5 |
| *Video retrieval* | | | | | | |
| Chamfer (DINOv2) | 31.9 | 35.4 | 35.8 | 32.4 | 33.2 | 34.0 |
| Sym. chamfer (DINOv2) | 30.0 | 31.0 | 32.7 | 30.3 | 30.3 | 31.0 |
| ViSiL‡ | 30.5 | – | 30.7 | – | – | – |
| *SupCon head (trained)* | | | | | | |
| C3D | 42.2 | 48.8 | 53.4 | 43.8 | 44.4 | 45.2 |
| DINOv2 | 63.1 | 71.8 | 74.4 | 64.5 | 66.0 | 67.8 |
| VideoMAE V2 | **66.7** | **75.3** | 78.5 | **69.5** | 69.5 | 70.7 |
| V-JEPA 2 | 65.4 | 73.8 | **78.5** | 68.1 | **69.7** | **70.7** |
| PeskaVLP | 64.0 | 74.8 | 77.2 | 66.3 | 67.6 | 69.3 |
| SurgRec-MAE | 48.8 | 54.8 | 57.6 | 50.8 | 51.1 | 52.0 |
| *Classifier oracle: hidden features* | | | | | | |
| C3D | 36.1 | 37.8 | 41.4 | 36.3 | 36.9 | 37.1 |
| DINOv2 | 49.6 | 55.5 | 58.9 | 50.6 | 51.9 | 53.0 |
| VideoMAE V2 | 48.1 | 53.6 | 56.2 | 49.0 | 49.8 | 50.8 |
| V-JEPA 2 | 50.6 | 56.4 | 62.6 | 51.8 | 52.8 | 53.8 |
| PeskaVLP | 55.8 | 63.8 | 65.8 | 56.6 | 57.7 | 59.4 |
| SurgRec-MAE | 35.3 | 38.2 | 40.0 | 36.3 | 36.7 | 37.0 |
| *Classifier oracle: predicted label* | | | | | | |
| C3D | 34.3 | 37.8 | 43.1 | 35.2 | 35.7 | 36.2 |
| DINOv2 | 49.3 | 57.1 | 60.8 | 50.5 | 51.8 | 52.9 |
| VideoMAE V2 | 49.9 | 64.3 | 67.2 | 53.2 | 54.7 | 54.9 |
| V-JEPA 2 | 49.5 | 59.6 | 62.6 | 51.2 | 52.3 | 52.3 |
| PeskaVLP | 52.3 | 61.8 | 64.4 | 53.4 | 54.7 | 55.4 |
| SurgRec-MAE | 38.7 | 45.8 | 49.2 | 41.5 | 41.2 | 40.7 |

Table 8. SCENE retrieval mAP (%), Prostatectomy, within-dataset galleries, per granularity. Speed-$s$ queries are searched in the short gallery. Short-clip queries: 940 from 3 test cases. †Ranks by the duration of the source segment, which the clips do not reveal. ‡500 random queries per matrix. Bold: best non-reference method per column.

| Method | Short | Medium | Phase | 2× | 3× | 5× |
|---|---|---|---|---|---|---|
| *Reference* | | | | | | |
| Random (chance) | 13.8 | 13.9 | 13.1 | 14.0 | 13.9 | 13.2 |
| Duration only† | 17.0 | 17.5 | 16.4 | 17.0 | 17.1 | 16.3 |
| *Frozen encoders (mean pooling)* | | | | | | |
| C3D | 25.2 | 26.3 | 25.0 | 24.6 | 24.0 | 22.1 |
| DINOv2 | 31.9 | 33.8 | 33.0 | 31.9 | 32.2 | 31.4 |
| VideoMAE V2 | 31.2 | 33.1 | 32.9 | 31.6 | 31.7 | 30.5 |
| V-JEPA 2 | 36.0 | 38.8 | 35.7 | 36.0 | 34.2 | 30.9 |
| PeskaVLP | 43.0 | 45.8 | 44.6 | 44.1 | 44.6 | 43.0 |
| SurgRec-MAE | 23.5 | 24.8 | 24.5 | 23.1 | 22.6 | 21.2 |
| *Video retrieval* | | | | | | |
| Chamfer (DINOv2) | 32.5 | 34.1 | 33.5 | 32.8 | 33.0 | 32.1 |
| Sym. chamfer (DINOv2) | 30.8 | 31.9 | 31.2 | 30.7 | 30.8 | 30.1 |
| ViSiL‡ | 31.0 | – | 32.0 | – | – | – |
| *SupCon head (trained)* | | | | | | |
| C3D | 44.4 | 46.6 | 46.2 | 45.2 | 45.1 | 43.9 |
| DINOv2 | 64.6 | 67.4 | 69.0 | 66.4 | 67.8 | 67.8 |
| VideoMAE V2 | 64.6 | 67.9 | 69.2 | 66.9 | 68.2 | 68.5 |
| V-JEPA 2 | **68.1** | **72.1** | **74.3** | **69.3** | **71.5** | **72.9** |
| PeskaVLP | 62.5 | 65.2 | 66.8 | 64.7 | 65.3 | 64.7 |
| SurgRec-MAE | 61.7 | 64.5 | 67.3 | 63.6 | 65.7 | 64.8 |
| *Classifier oracle: hidden features* | | | | | | |
| C3D | 32.0 | 33.6 | 32.6 | 32.0 | 31.6 | 29.4 |
| DINOv2 | 51.9 | 54.5 | 54.1 | 52.4 | 53.4 | 53.2 |
| VideoMAE V2 | 47.7 | 51.2 | 51.5 | 50.1 | 50.0 | 47.8 |
| V-JEPA 2 | 56.9 | 61.7 | 62.9 | 58.9 | 59.5 | 59.5 |
| PeskaVLP | 53.1 | 55.6 | 54.2 | 54.0 | 55.1 | 53.3 |
| SurgRec-MAE | 42.4 | 45.8 | 47.8 | 43.8 | 45.0 | 44.7 |
| *Classifier oracle: predicted label* | | | | | | |
| C3D | 28.5 | 30.0 | 29.3 | 29.2 | 29.3 | 28.1 |
| DINOv2 | 46.2 | 51.0 | 54.5 | 47.8 | 49.1 | 50.2 |
| VideoMAE V2 | 53.0 | 56.9 | 57.6 | 54.1 | 54.5 | 53.2 |
| V-JEPA 2 | 54.6 | 61.2 | 65.0 | 57.8 | 57.6 | 58.2 |
| PeskaVLP | 48.5 | 50.6 | 53.1 | 49.1 | 51.1 | 50.8 |
| SurgRec-MAE | 39.8 | 46.8 | 45.1 | 42.7 | 42.5 | 41.0 |

Table 9. SCENE retrieval mAP (%), Robotic training (SurgVU), within-dataset galleries, per granularity. Speed-$s$ queries are searched in the short gallery. Short-clip queries: 1,855 from 58 test cases. †Ranks by the duration of the source segment, which the clips do not reveal. ‡500 random queries per matrix. Bold: best non-reference method per column.

| Method | Short | Medium | Phase | 2× | 3× | 5× |
|---|---|---|---|---|---|---|
| *Reference* | | | | | | |
| Random (chance) | 33.9 | 34.4 | 33.9 | 34.0 | 33.9 | 33.9 |
| Duration only† | 38.9 | 39.3 | 38.9 | 38.9 | 38.9 | 38.9 |
| *Frozen encoders (mean pooling)* | | | | | | |
| C3D | 40.9 | 44.5 | 45.2 | 40.7 | 40.8 | 40.7 |
| DINOv2 | 48.3 | 52.8 | 53.7 | 48.2 | 48.4 | 48.8 |
| VideoMAE V2 | 48.1 | 54.1 | 56.2 | 48.8 | 49.0 | 49.3 |
| V-JEPA 2 | 44.7 | 50.7 | 54.2 | 44.7 | 44.4 | 44.1 |
| PeskaVLP | 52.4 | 55.5 | 56.4 | 52.2 | 52.2 | 52.4 |
| SurgRec-MAE | 43.5 | 46.4 | 48.9 | 42.9 | 42.3 | 41.5 |
| *Video retrieval* | | | | | | |
| Chamfer (DINOv2) | 49.5 | 53.8 | 54.8 | 49.6 | 49.9 | 50.1 |
| Sym. chamfer (DINOv2) | 47.9 | 50.7 | 51.5 | 47.9 | 48.1 | 48.3 |
| ViSiL‡ | 50.4 | – | 49.1 | – | – | – |
| *SupCon head (trained)* | | | | | | |
| C3D | 76.1 | 83.1 | 84.9 | 77.2 | 77.4 | 78.8 |
| DINOv2 | 91.3 | 94.0 | 95.2 | **92.8** | **92.7** | 92.9 |
| VideoMAE V2 | 90.4 | 94.0 | 95.1 | 90.7 | 91.3 | 91.7 |
| V-JEPA 2 | **91.3** | **95.1** | **95.9** | 91.8 | 92.4 | **92.9** |
| PeskaVLP | 89.1 | 92.0 | 93.5 | 89.5 | 89.8 | 90.3 |
| SurgRec-MAE | 85.4 | 90.0 | 90.6 | 83.9 | 85.7 | 86.5 |
| *Classifier oracle: hidden features* | | | | | | |
| C3D | 56.0 | 62.7 | 64.1 | 56.3 | 56.8 | 57.4 |
| DINOv2 | 77.9 | 82.0 | 83.3 | 78.3 | 78.8 | 79.1 |
| VideoMAE V2 | 70.4 | 77.5 | 79.4 | 71.4 | 71.7 | 72.2 |
| V-JEPA 2 | 77.6 | 84.4 | 86.3 | 78.2 | 78.3 | 79.0 |
| PeskaVLP | 72.0 | 76.7 | 78.1 | 72.5 | 72.7 | 73.3 |
| SurgRec-MAE | 64.2 | 73.6 | 74.3 | 63.1 | 63.9 | 65.1 |
| *Classifier oracle: predicted label* | | | | | | |
| C3D | 67.5 | 73.4 | 76.6 | 68.6 | 68.4 | 68.6 |
| DINOv2 | 82.9 | 86.1 | 88.1 | 83.9 | 84.1 | 83.5 |
| VideoMAE V2 | 81.0 | 88.0 | 89.2 | 81.1 | 81.4 | 82.0 |
| V-JEPA 2 | 82.5 | 89.5 | 90.6 | 83.1 | 84.1 | 84.5 |
| PeskaVLP | 81.2 | 84.3 | 86.7 | 81.7 | 81.9 | 82.3 |
| SurgRec-MAE | 75.8 | 83.7 | 85.0 | 74.6 | 76.6 | 77.0 |

Table 10. SCENE retrieval mAP (%), Colonoscopy, within-dataset galleries, per granularity. Speed-$s$ queries are searched in the short gallery. Short-clip queries: 145 from 11 test cases. †Ranks by the duration of the source segment, which the clips do not reveal. ‡500 random queries per matrix. Bold: best non-reference method per column.

| Method | Short | 2× | 3× |
|---|---|---|---|
| *Reference* | | | |
| Random (chance) | 45.5 | 48.2 | 48.4 |
| Duration only† | 51.5 | 48.4 | 51.4 |
| *Frozen encoders (mean pooling)* | | | |
| C3D | **60.5** | **63.1** | 56.6 |
| DINOv2 | 48.9 | 50.6 | 43.6 |
| VideoMAE V2 | 51.4 | 50.8 | 45.0 |
| V-JEPA 2 | 54.9 | 54.9 | 57.5 |
| PeskaVLP | 54.9 | 56.7 | **58.1** |
| SurgRec-MAE | 54.4 | 53.2 | 52.2 |
| *Video retrieval* | | | |
| Chamfer (DINOv2) | 48.8 | 49.6 | 44.0 |
| Sym. chamfer (DINOv2) | 49.4 | 53.0 | 45.6 |
| ViSiL‡ | 55.2 | – | – |
| *SupCon head (trained)* | | | |
| C3D | 46.2 | 51.1 | 45.3 |
| DINOv2 | 43.5 | 46.0 | 43.8 |
| VideoMAE V2 | 47.2 | 46.9 | 44.5 |
| V-JEPA 2 | 51.3 | 51.9 | 54.1 |
| PeskaVLP | 50.5 | 52.2 | 51.0 |
| SurgRec-MAE | 51.9 | 52.7 | 49.1 |
| *Classifier oracle: hidden features* | | | |
| C3D | 48.2 | 52.7 | 46.0 |
| DINOv2 | 48.2 | 51.8 | 42.4 |
| VideoMAE V2 | 49.2 | 49.6 | 44.7 |
| V-JEPA 2 | 51.0 | 54.5 | 54.6 |
| PeskaVLP | 55.0 | 57.2 | 57.0 |
| SurgRec-MAE | 53.9 | 49.9 | 53.1 |
| *Classifier oracle: predicted label* | | | |
| C3D | 44.8 | 49.0 | 47.3 |
| DINOv2 | 45.9 | 49.2 | 49.4 |
| VideoMAE V2 | 45.5 | 48.6 | 48.2 |
| V-JEPA 2 | 44.8 | 47.0 | 48.0 |
| PeskaVLP | 46.7 | 51.1 | 48.5 |
| SurgRec-MAE | 47.9 | 51.5 | 48.0 |

Table 11. Effect of temporal granularity and playback speed (SCENE, within dataset, mAP %). Left: averaged over the 5 procedures with phase and speed clips (Phaco, Gonio, Chole, Prost, SurgVU); speed-$s$ queries are searched in the short gallery. Right: averaged over the 3 procedures with medium clips (Chole, Prost, SurgVU). –: not computed for this method. †Ranks by the duration of the source segment, which the clips do not reveal. ‡500 random queries per matrix. Bold: best non-reference method per column.

| | 5 procedures | | | | | 3 procedures | | |
|---|---|---|---|---|---|---|---|---|
| Method | Short | Phase | 2× | 3× | 5× | Short | Medium | Phase |
| *Reference* | | | | | | | | |
| Random (chance) | 18.5 | 20.6 | 18.5 | 18.5 | 18.6 | 21.9 | 22.2 | 21.6 |
| Duration only† | 26.2 | 27.9 | 26.3 | 26.5 | 27.3 | 28.8 | 29.0 | 28.5 |
| *Frozen encoders (mean pooling)* | | | | | | | | |
| C3D | 26.9 | 33.6 | 26.9 | 27.2 | 27.5 | 31.5 | 33.6 | 34.8 |
| DINOv2 | 32.1 | 38.4 | 32.2 | 32.8 | 33.5 | 37.2 | 40.7 | 40.9 |
| VideoMAE V2 | 33.4 | 40.7 | 33.7 | 34.0 | 34.2 | 36.6 | 40.4 | 41.6 |
| V-JEPA 2 | 31.5 | 39.6 | 31.6 | 31.6 | 31.4 | 37.6 | 41.9 | 43.4 |
| PeskaVLP | 38.4 | 44.7 | 38.7 | 39.3 | 39.8 | 47.1 | 51.3 | 51.5 |
| SurgRec-MAE | 26.4 | 31.3 | 26.3 | 26.3 | 26.0 | 30.9 | 32.8 | 33.5 |
| *Video retrieval* | | | | | | | | |
| Chamfer (DINOv2) | 33.0 | 39.2 | 33.2 | 33.9 | 34.7 | 38.0 | 41.1 | 41.4 |
| Sym. chamfer (DINOv2) | 31.6 | 36.7 | 31.7 | 32.1 | 32.8 | 36.2 | 37.9 | 38.5 |
| ViSiL‡ | 34.5 | 38.5 | – | – | – | 37.3 | – | 37.3 |
| *SupCon head (trained)* | | | | | | | | |
| C3D | 47.2 | 56.9 | 48.5 | 49.5 | 51.3 | 54.2 | 59.5 | 61.5 |
| DINOv2 | 65.2 | 73.6 | 66.6 | 68.2 | 70.1 | 73.0 | 77.7 | 79.5 |
| VideoMAE V2 | **70.9** | 78.0 | **72.5** | **73.6** | **75.3** | 73.9 | 79.1 | 80.9 |
| V-JEPA 2 | 69.5 | **78.2** | 71.0 | 72.8 | 74.8 | **74.9** | **80.3** | **82.9** |
| PeskaVLP | 63.4 | 71.6 | 64.8 | 66.1 | 67.9 | 71.9 | 77.3 | 79.2 |
| SurgRec-MAE | 58.3 | 67.1 | 59.2 | 60.8 | 62.8 | 65.3 | 69.8 | 71.9 |
| *Classifier oracle: hidden features* | | | | | | | | |
| C3D | 35.9 | 44.0 | 36.4 | 37.1 | 37.9 | 41.4 | 44.7 | 46.1 |
| DINOv2 | 51.5 | 59.1 | 52.0 | 53.2 | 54.4 | 59.8 | 64.0 | 65.4 |
| VideoMAE V2 | 52.4 | 60.7 | 53.7 | 54.4 | 54.7 | 55.4 | 60.8 | 62.4 |
| V-JEPA 2 | 52.5 | 63.1 | 53.8 | 54.9 | 56.4 | 61.7 | 67.5 | 70.6 |
| PeskaVLP | 49.8 | 56.9 | 50.5 | 51.3 | 52.1 | 60.3 | 65.4 | 66.0 |
| SurgRec-MAE | 41.9 | 50.6 | 42.6 | 43.4 | 44.1 | 47.3 | 52.5 | 54.0 |
| *Classifier oracle: predicted label* | | | | | | | | |
| C3D | 37.8 | 46.8 | 38.7 | 39.6 | 40.6 | 43.4 | 47.1 | 49.7 |
| DINOv2 | 52.6 | 62.0 | 53.5 | 54.9 | 56.5 | 59.5 | 64.7 | 67.8 |
| VideoMAE V2 | 58.5 | 67.9 | 59.9 | 61.1 | 62.0 | 61.3 | 69.8 | 71.4 |
| V-JEPA 2 | 56.3 | 66.7 | 57.8 | 58.8 | 60.0 | 62.2 | 70.1 | 72.7 |
| PeskaVLP | 51.7 | 59.3 | 52.1 | 53.2 | 54.5 | 60.7 | 65.6 | 68.1 |
| SurgRec-MAE | 45.3 | 55.7 | 46.6 | 47.4 | 48.0 | 51.4 | 58.8 | 59.8 |

Table 12. SCENE retrieval, short clips, within dataset: all metrics (%), averaged over the six procedures (ViSiL: 500 queries per procedure). nDCG@10 uses graded gains (2 same phase, 1 neighbouring phase). †Ranks by the duration of the source segment, which the clips do not reveal. ‡500 random queries per matrix. Bold: best non-reference method per column.

| Method | mAP | $mAP_{macro}$ | R@1 | R@5 | R@10 | P@10 | nDCG@10 |
|---|---|---|---|---|---|---|---|
| *Reference* | | | | | | | |
| Random (chance) | 23.0 | 20.5 | 22.0 | 63.3 | 81.2 | 21.9 | 31.2 |
| Duration only† | 30.5 | 28.2 | 31.0 | 32.8 | 47.2 | 31.1 | 39.0 |
| *Frozen encoders (mean pooling)* | | | | | | | |
| C3D | 32.5 | 29.7 | 48.4 | 73.7 | 83.4 | 42.9 | 51.3 |
| DINOv2 | 34.9 | 32.3 | 54.3 | 79.3 | 88.8 | 49.4 | 57.9 |
| VideoMAE V2 | 36.4 | 33.4 | 57.7 | 83.2 | **92.6** | 52.2 | 60.1 |
| V-JEPA 2 | 35.4 | 32.3 | 57.0 | 79.4 | 87.9 | 49.9 | 58.8 |
| PeskaVLP | 41.1 | 37.6 | 57.8 | 79.2 | 87.8 | 55.1 | 62.7 |
| SurgRec-MAE | 31.1 | 28.1 | 45.5 | 74.3 | 84.4 | 41.9 | 50.6 |
| *Video retrieval* | | | | | | | |
| Chamfer (DINOv2) | 35.6 | 33.0 | 54.7 | 81.1 | 90.1 | 50.4 | 58.9 |
| Sym. chamfer (DINOv2) | 34.5 | 32.4 | 50.7 | 78.5 | 88.5 | 48.3 | 56.9 |
| ViSiL‡ | 37.9 | 34.5 | 57.3 | 82.9 | 90.2 | 53.7 | 61.5 |
| *SupCon head (trained)* | | | | | | | |
| C3D | 47.0 | 43.4 | 54.2 | 76.5 | 84.1 | 51.6 | 59.7 |
| DINOv2 | 61.6 | 58.8 | 67.9 | 82.5 | 88.0 | 65.3 | 72.7 |
| VideoMAE V2 | **67.0** | **64.4** | 71.9 | **86.0** | 91.4 | 70.6 | 76.9 |
| V-JEPA 2 | 66.5 | 63.9 | **73.0** | 85.4 | 90.4 | **70.9** | **77.1** |
| PeskaVLP | 61.2 | 58.6 | 66.4 | 79.8 | 87.9 | 64.9 | 71.9 |
| SurgRec-MAE | 57.2 | 54.3 | 63.6 | 81.6 | 87.5 | 62.6 | 69.4 |
| *Classifier oracle: hidden features* | | | | | | | |
| C3D | 38.0 | 35.4 | 49.6 | 72.8 | 83.4 | 47.4 | 55.6 |
| DINOv2 | 50.9 | 48.6 | 62.9 | 80.3 | 89.1 | 60.6 | 68.0 |
| VideoMAE V2 | 51.9 | 49.6 | 67.6 | 84.7 | 90.9 | 63.3 | 70.5 |
| V-JEPA 2 | 52.2 | 50.3 | 66.0 | 82.5 | 88.9 | 63.1 | 70.4 |
| PeskaVLP | 50.7 | 47.7 | 61.9 | 80.5 | 89.1 | 61.4 | 68.2 |
| SurgRec-MAE | 43.9 | 41.3 | 58.7 | 80.0 | 87.9 | 56.5 | 64.0 |
| *Classifier oracle: predicted label* | | | | | | | |
| C3D | 39.0 | 35.8 | 46.9 | 74.5 | 82.4 | 42.7 | 51.2 |
| DINOv2 | 51.4 | 48.0 | 56.8 | 85.6 | 89.2 | 55.4 | 62.7 |
| VideoMAE V2 | 56.3 | 53.5 | 65.6 | 84.9 | 90.6 | 61.3 | 68.8 |
| V-JEPA 2 | 54.4 | 51.5 | 56.4 | 81.0 | 90.4 | 56.3 | 63.8 |
| PeskaVLP | 50.9 | 47.7 | 55.6 | 78.8 | 89.6 | 55.2 | 62.8 |
| SurgRec-MAE | 45.7 | 42.8 | 54.8 | 79.6 | 86.0 | 50.9 | 58.6 |

Table 13. Mean vs. max temporal pooling of frozen encoders (mAP %, within dataset): SCENE short (six procedures) and phase (five), WORKFLOW (phaco, goniotomy).

| | Short | | Phase | | Workflow | |
|---|---|---|---|---|---|---|
| Encoder | mean | max | mean | max | mean | max |
| C3D | 32.5 | 32.1 | 33.6 | 33.5 | 20.7 | 21.4 |
| DINOv2 | 34.9 | 34.6 | 38.4 | 37.6 | 21.4 | 22.2 |
| VideoMAE V2 | 36.4 | 33.2 | 40.7 | 36.4 | 23.9 | 21.7 |
| V-JEPA 2 | 35.4 | 34.6 | 39.6 | 39.0 | 21.7 | 22.9 |
| PeskaVLP | 41.1 | 40.7 | 44.7 | 43.9 | 21.9 | 24.5 |
| SurgRec-MAE | 31.1 | 27.9 | 31.3 | 28.0 | 20.9 | 20.9 |

Table 14. Cross-dataset SCENE retrieval, short clips (mAP %); column groups give the query dataset. W: gallery from the query's dataset; X: gallery from the other dataset (shared labels only); P: pooled gallery of both datasets. The classifier oracle and the SupCon head are trained on the query's dataset only. †Ranks by the duration of the source segment, which the clips do not reveal. ‡500 random queries per matrix. Bold: best non-reference method per column.

| | Phacoemulsification | | | | | | Cholecystectomy | | | | | |
|---|---|---|---|---|---|---|---|---|---|---|---|---|
| | Cat.-LMM | | | Cat.-1K | | | Cholec80 | | | HeiChole | | |
| Method | W | X | P | W | X | P | W | X | P | W | X | P |
| *Reference* | | | | | | | | | | | | |
| Random (chance) | 10.5 | 9.9 | 10.2 | 13.1 | 9.4 | 9.7 | 17.4 | 18.0 | 17.2 | 19.3 | 17.2 | 17.4 |
| Duration only† | 16.6 | 19.7 | 16.4 | 27.9 | 14.6 | 15.9 | 31.2 | 29.2 | 29.9 | 28.6 | 28.9 | 28.0 |
| *Frozen encoders (mean pooling)* | | | | | | | | | | | | |
| C3D | 13.0 | 12.0 | 12.5 | 22.8 | 10.7 | 14.1 | 29.0 | 25.7 | 27.6 | 26.4 | 24.7 | 24.4 |
| DINOv2 | 14.3 | 12.4 | 13.8 | 24.3 | 11.1 | 14.7 | 32.6 | 27.7 | 31.1 | 28.1 | 28.1 | 27.5 |
| VideoMAE V2 | 19.7 | 15.2 | 18.3 | 32.6 | 10.6 | 16.8 | 31.4 | 26.1 | 29.3 | 27.5 | 27.5 | 26.7 |
| V-JEPA 2 | 13.0 | 12.4 | 12.7 | 27.5 | 10.2 | 15.2 | 32.7 | 29.2 | 31.1 | 30.6 | 30.3 | 29.7 |
| PeskaVLP | 14.1 | 12.9 | 13.6 | 32.5 | 11.4 | 16.9 | 48.9 | 40.0 | 45.9 | 37.4 | 40.1 | 38.5 |
| SurgRec-MAE | 13.0 | 11.8 | 12.8 | 17.9 | 9.4 | 12.4 | 25.5 | 24.2 | 24.5 | 26.1 | 24.6 | 24.5 |
| *Video retrieval* | | | | | | | | | | | | |
| Chamfer (DINOv2) | 14.6 | 12.5 | 14.1 | 26.6 | 11.1 | 15.3 | 33.1 | 28.2 | 31.5 | 28.5 | 27.9 | 27.4 |
| Sym. chamfer (DINOv2) | 14.4 | 12.7 | 13.9 | 25.7 | 11.3 | 14.9 | 30.4 | 26.4 | 29.0 | 28.8 | 27.4 | 27.1 |
| ViSiL‡ | 17.9 | 17.4 | 16.9 | 34.0 | 15.6 | 18.4 | 31.2 | 27.2 | 29.6 | 28.6 | 28.2 | 27.6 |
| *SupCon head (trained)* | | | | | | | | | | | | |
| C3D | 22.5 | 13.0 | 20.3 | 42.2 | 9.9 | 16.4 | 45.8 | 33.0 | 42.1 | 32.2 | 29.7 | 29.8 |
| DINOv2 | 41.3 | 29.6 | 38.5 | 73.6 | 19.3 | 35.3 | 69.9 | 50.6 | 64.6 | 44.0 | 42.9 | 42.6 |
| VideoMAE V2 | **60.8** | 35.8 | **55.7** | **82.0** | **24.7** | **42.2** | **72.9** | **56.9** | **68.8** | **49.1** | 45.5 | 45.9 |
| V-JEPA 2 | 52.9 | **37.3** | 49.8 | 80.2 | 22.3 | 39.5 | 71.8 | 51.8 | 66.4 | 47.5 | **48.7** | **47.7** |
| PeskaVLP | 34.0 | 19.7 | 31.5 | 71.2 | 13.1 | 27.7 | 71.1 | 51.3 | 66.0 | 44.3 | 45.6 | 44.9 |
| SurgRec-MAE | 36.6 | 16.2 | 33.2 | 51.9 | 13.7 | 23.9 | 51.2 | 41.1 | 47.4 | 42.0 | 33.5 | 34.6 |
| *Classifier oracle: hidden features* | | | | | | | | | | | | |
| C3D | 15.5 | 12.9 | 14.8 | 32.0 | 10.8 | 16.2 | 36.9 | 31.2 | 34.8 | 33.9 | 30.9 | 31.1 |
| DINOv2 | 24.5 | 21.8 | 23.2 | 49.1 | 15.7 | 23.8 | 53.2 | 43.6 | 50.3 | 39.5 | 38.1 | 37.8 |
| VideoMAE V2 | 35.2 | 29.0 | 32.1 | 58.0 | 19.4 | 30.9 | 51.2 | 40.3 | 47.9 | 39.3 | 37.1 | 36.9 |
| V-JEPA 2 | 22.5 | 20.5 | 21.1 | 57.5 | 18.2 | 28.2 | 53.7 | 42.6 | 50.5 | 42.0 | 41.8 | 41.1 |
| PeskaVLP | 18.5 | 16.6 | 17.7 | 49.2 | 13.6 | 22.1 | 60.0 | 47.3 | 56.4 | 43.9 | 47.4 | 45.8 |
| SurgRec-MAE | 21.9 | 15.3 | 20.8 | 33.1 | 13.3 | 18.9 | 35.9 | 35.1 | 34.4 | 33.6 | 33.2 | 32.6 |
| *Classifier oracle: predicted label* | | | | | | | | | | | | |
| C3D | 17.3 | 11.7 | 15.6 | 28.1 | 9.7 | 12.0 | 36.2 | 27.5 | 33.7 | 28.8 | 25.8 | 26.0 |
| DINOv2 | 33.2 | 19.9 | 30.2 | 56.2 | 13.4 | 19.4 | 53.7 | 39.9 | 49.5 | 37.0 | 34.1 | 34.3 |
| VideoMAE V2 | 50.2 | 15.4 | 40.2 | 64.2 | 13.1 | 21.3 | 54.9 | 39.7 | 49.9 | 35.7 | 33.6 | 33.7 |
| V-JEPA 2 | 40.0 | 17.7 | 34.9 | 63.2 | 14.7 | 23.1 | 53.6 | 40.6 | 49.4 | 38.0 | 36.1 | 36.1 |
| PeskaVLP | 22.7 | 12.3 | 21.1 | 51.9 | 10.7 | 15.8 | 56.9 | 37.8 | 51.1 | 39.7 | 37.0 | 37.2 |
| SurgRec-MAE | 28.7 | 12.5 | 24.9 | 30.4 | 10.3 | 15.7 | 41.5 | 34.2 | 38.9 | 30.7 | 25.7 | 26.5 |

Table 15. Cross-dataset SCENE retrieval, phase clips (mAP %); column groups give the query dataset. W: gallery from the query's dataset; X: gallery from the other dataset (shared labels only); P: pooled gallery of both datasets. The classifier oracle and the SupCon head are trained on the query's dataset only. †Ranks by the duration of the source segment, which the clips do not reveal. ‡500 random queries per matrix. Bold: best non-reference method per column.

| | Phacoemulsification | | | | | | Cholecystectomy | | | | | |
|---|---|---|---|---|---|---|---|---|---|---|---|---|
| | Cat.-LMM | | | Cat.-1K | | | Cholec80 | | | HeiChole | | |
| Method | W | X | P | W | X | P | W | X | P | W | X | P |
| *Reference* | | | | | | | | | | | | |
| Random (chance) | 15.3 | 20.7 | 15.6 | 22.2 | 19.9 | 18.3 | 17.3 | 18.2 | 17.2 | 19.6 | 17.4 | 17.6 |
| Duration only† | 21.6 | 34.0 | 22.6 | 39.4 | 25.9 | 25.5 | 31.1 | 28.8 | 29.6 | 27.5 | 28.4 | 27.3 |
| *Frozen encoders (mean pooling)* | | | | | | | | | | | | |
| C3D | 21.5 | 25.1 | 20.9 | 39.7 | 21.6 | 24.3 | 35.5 | 29.3 | 33.3 | 29.9 | 29.1 | 28.5 |
| DINOv2 | 21.4 | 23.8 | 20.5 | 35.6 | 24.3 | 24.2 | 37.4 | 30.0 | 35.3 | 31.9 | 31.0 | 30.6 |
| VideoMAE V2 | 27.6 | 26.9 | 25.2 | 47.4 | 23.0 | 25.2 | 37.5 | 28.7 | 34.5 | 31.0 | 31.8 | 30.8 |
| V-JEPA 2 | 21.3 | 23.7 | 20.3 | 44.0 | 24.9 | 25.7 | 42.2 | 34.0 | 39.4 | 34.6 | 37.9 | 36.4 |
| PeskaVLP | 21.1 | 25.2 | 20.3 | 40.3 | 21.3 | 23.6 | 57.1 | 44.7 | 52.9 | 43.0 | 45.4 | 43.5 |
| SurgRec-MAE | 19.7 | 25.7 | 18.9 | 33.0 | 24.0 | 22.7 | 26.9 | 25.0 | 25.7 | 27.1 | 26.1 | 25.7 |
| *Video retrieval* | | | | | | | | | | | | |
| Chamfer (DINOv2) | 22.0 | 24.4 | 21.0 | 37.8 | 24.1 | 24.4 | 37.3 | 30.1 | 35.2 | 31.6 | 29.9 | 29.5 |
| Sym. chamfer (DINOv2) | 21.6 | 23.9 | 20.6 | 37.0 | 24.4 | 24.4 | 33.3 | 27.5 | 31.3 | 30.9 | 29.7 | 29.2 |
| ViSiL‡ | 25.9 | 27.9 | 24.2 | 43.0 | 24.2 | 25.3 | 31.3 | 25.9 | 29.4 | 29.3 | 28.2 | 27.6 |
| *SupCon head (trained)* | | | | | | | | | | | | |
| C3D | 35.0 | 25.9 | 32.1 | 55.5 | 20.4 | 26.1 | 58.4 | 37.8 | 52.9 | 38.9 | 35.9 | 36.0 |
| DINOv2 | 51.3 | 37.3 | 46.8 | 82.6 | 32.8 | 42.5 | 82.2 | 60.7 | 76.7 | 52.4 | 50.8 | 50.5 |
| VideoMAE V2 | **66.4** | 32.4 | **61.2** | 86.6 | 27.6 | 41.0 | **85.4** | **70.2** | **81.7** | 58.6 | 54.5 | 54.8 |
| V-JEPA 2 | 59.9 | **49.1** | 56.9 | **87.6** | 34.3 | **44.5** | 84.7 | 62.5 | 79.4 | **61.0** | **60.1** | **59.1** |
| PeskaVLP | 44.0 | 33.4 | 41.2 | 74.6 | 21.2 | 32.1 | 84.5 | 59.5 | 78.5 | 56.8 | 56.7 | 56.2 |
| SurgRec-MAE | 47.6 | 32.2 | 43.1 | 65.6 | 25.9 | 34.1 | 60.0 | 46.1 | 55.6 | 50.9 | 39.0 | 40.6 |
| *Classifier oracle: hidden features* | | | | | | | | | | | | |
| C3D | 25.4 | 24.9 | 24.2 | 48.2 | 22.8 | 26.8 | 43.1 | 36.5 | 40.5 | 36.8 | 33.8 | 33.5 |
| DINOv2 | 33.2 | 30.6 | 31.4 | 58.5 | 28.9 | 32.3 | 63.4 | 50.6 | 59.6 | 46.0 | 42.1 | 42.1 |
| VideoMAE V2 | 41.7 | 33.1 | 38.1 | 69.7 | 31.0 | 36.3 | 60.0 | 47.7 | 56.3 | 45.8 | 42.3 | 42.2 |
| V-JEPA 2 | 32.9 | 32.5 | 31.1 | 72.2 | **37.0** | 40.3 | 66.9 | 53.8 | 63.1 | 50.2 | 51.4 | 50.1 |
| PeskaVLP | 26.3 | 27.4 | 25.1 | 53.1 | 24.6 | 28.6 | 71.0 | 53.8 | 66.2 | 50.9 | 54.4 | 52.5 |
| SurgRec-MAE | 30.2 | 30.0 | 28.7 | 49.6 | 33.3 | 34.5 | 40.4 | 39.3 | 38.7 | 38.7 | 38.2 | 37.4 |
| *Classifier oracle: predicted label* | | | | | | | | | | | | |
| C3D | 28.5 | 21.8 | 25.9 | 40.3 | 19.5 | 20.7 | 46.7 | 32.1 | 42.2 | 33.0 | 27.9 | 28.6 |
| DINOv2 | 43.2 | 27.0 | 38.6 | 70.1 | 24.2 | 27.1 | 66.1 | 47.6 | 60.6 | 45.8 | 41.9 | 42.2 |
| VideoMAE V2 | 56.6 | 21.8 | 46.4 | 70.5 | 21.8 | 30.0 | 75.1 | 54.4 | 68.3 | 45.0 | 39.7 | 40.2 |
| V-JEPA 2 | 46.7 | 29.8 | 42.4 | 68.0 | 26.8 | 32.6 | 68.6 | 50.4 | 62.9 | 45.6 | 42.5 | 42.5 |
| PeskaVLP | 29.8 | 21.3 | 27.2 | 53.4 | 21.3 | 23.0 | 70.8 | 42.6 | 62.0 | 46.1 | 45.6 | 45.0 |
| SurgRec-MAE | 39.7 | 25.0 | 35.4 | 46.7 | 25.2 | 27.3 | 52.5 | 40.6 | 48.8 | 40.0 | 30.0 | 31.8 |

Table 16. WORKFLOW (transition) retrieval (%). W/X/P: within-dataset, cross-dataset and pooled galleries; R@1 and nDCG@10 within dataset. nDCG gains: 2 same transition type, 1 same phase before or after. †Ranks by the duration of the source segment, which the clips do not reveal. ‡500 random queries per matrix. Bold: best non-reference method per column.

| | Phacoemulsification | | | | | Goniotomy | | |
|---|---|---|---|---|---|---|---|---|
| Method | mAP W | mAP X | mAP P | R@1 | nDCG | mAP | R@1 | nDCG |
| *Reference* | | | | | | | | |
| Random (chance) | 9.8 | 14.2 | 8.3 | 10.0 | 11.3 | 21.4 | 19.2 | 24.8 |
| Duration only† | 12.4 | 14.3 | 9.3 | 12.1 | 14.4 | 25.4 | 26.3 | 29.5 |
| *Frozen encoders (mean pooling)* | | | | | | | | |
| C3D | 12.7 | 15.9 | 10.7 | 20.7 | 19.4 | 28.7 | 44.4 | 38.8 |
| DINOv2 | 12.6 | 15.6 | 10.7 | 18.3 | 18.5 | 30.2 | 37.4 | 37.4 |
| VideoMAE V2 | 16.3 | 15.8 | 12.8 | 28.1 | 24.7 | 31.4 | 44.4 | 41.4 |
| V-JEPA 2 | 13.9 | 13.9 | 11.3 | 21.0 | 20.0 | 29.4 | 33.3 | 35.0 |
| PeskaVLP | 14.8 | 17.1 | 11.8 | 23.8 | 21.2 | 29.0 | 33.3 | 35.5 |
| SurgRec-MAE | 11.7 | 14.9 | 10.2 | 17.3 | 16.8 | 30.1 | 41.4 | 37.5 |
| *Video retrieval* | | | | | | | | |
| Chamfer (DINOv2) | 13.3 | 15.5 | 11.2 | 20.0 | 19.9 | 31.6 | 40.4 | 39.5 |
| Sym. chamfer (DINOv2) | 12.9 | 15.9 | 10.8 | 18.8 | 18.6 | 30.1 | 39.4 | 38.3 |
| ViSiL‡ | 19.1 | 20.8 | 15.1 | 27.8 | 28.6 | 45.7 | 58.6 | 57.9 |
| *SupCon head (trained)* | | | | | | | | |
| C3D | 19.0 | 18.1 | 14.1 | 24.5 | 25.7 | 31.4 | 35.4 | 37.4 |
| DINOv2 | 27.1 | 23.7 | 20.4 | 35.3 | 35.7 | 38.4 | 39.4 | 47.9 |
| VideoMAE V2 | **40.7** | **29.4** | **29.4** | **49.9** | **50.3** | **53.9** | 52.5 | **62.6** |
| V-JEPA 2 | 37.2 | 27.6 | 26.8 | 47.2 | 47.3 | 52.8 | 57.6 | 59.8 |
| PeskaVLP | 25.8 | 20.3 | 19.4 | 33.1 | 34.2 | 39.3 | 51.5 | 48.6 |
| SurgRec-MAE | 23.4 | 17.3 | 17.6 | 32.6 | 31.3 | 50.6 | **60.6** | 57.3 |
| *Classifier oracle: hidden features* | | | | | | | | |
| C3D | 14.8 | 16.6 | 12.0 | 20.8 | 20.5 | 26.3 | 38.4 | 34.2 |
| DINOv2 | 16.1 | 19.4 | 12.5 | 22.3 | 22.6 | 33.4 | 43.4 | 42.1 |
| VideoMAE V2 | 23.2 | 20.6 | 17.1 | 34.8 | 31.5 | 32.0 | 47.5 | 43.0 |
| V-JEPA 2 | 20.1 | 17.5 | 14.1 | 24.2 | 26.1 | 30.7 | 37.4 | 35.9 |
| PeskaVLP | 19.3 | 20.4 | 14.4 | 25.3 | 25.3 | 31.2 | 37.4 | 39.0 |
| SurgRec-MAE | 14.5 | 15.8 | 12.0 | 22.7 | 21.7 | 38.4 | 57.6 | 49.4 |
| *Classifier oracle: predicted label* | | | | | | | | |
| C3D | 15.4 | 14.9 | 11.1 | 15.7 | 21.9 | 24.8 | 26.3 | 30.7 |
| DINOv2 | 19.7 | 17.2 | 14.5 | 25.3 | 28.7 | 37.9 | 47.5 | 46.8 |
| VideoMAE V2 | 27.8 | 16.9 | 16.7 | 28.0 | 34.3 | 30.0 | 34.3 | 37.7 |
| V-JEPA 2 | 24.4 | 14.5 | 13.2 | 25.6 | 30.2 | 31.9 | 39.4 | 38.0 |
| PeskaVLP | 20.6 | 15.9 | 13.4 | 23.5 | 27.7 | 32.9 | 25.2 | 39.5 |
| SurgRec-MAE | 16.1 | 14.0 | 11.8 | 17.8 | 21.8 | 39.2 | 33.3 | 45.4 |

Table 17. SCENE short-clip mAP (%) with 95% bootstrap confidence intervals over test cases (1,000 resamples), within dataset. Number of test cases in brackets in the header; intervals over very few cases (prostatectomy) are unreliable.

| Method | Phaco (23) | Gonio. (59) | Chole. (30) | Prost. (3) | SurgVU (58) | Colon (11) |
|---|---|---|---|---|---|---|
| Random | 11.0 [10.7, 11.2] | 15.7 [15.5, 16.0] | 17.9 [17.5, 18.3] | 13.8 [11.4, 14.7] | 33.9 [33.2, 34.5] | 45.5 [43.0, 48.9] |
| Duration only | 18.7 [17.7, 19.8] | 26.1 [25.1, 27.2] | 30.6 [29.3, 31.9] | 17.0 [15.2, 18.0] | 38.9 [37.8, 39.9] | 51.5 [44.7, 59.8] |
| C3D | 14.8 [14.0, 15.6] | 25.2 [24.1, 26.3] | 28.4 [27.6, 29.2] | 25.2 [21.5, 26.4] | 40.9 [40.0, 41.9] | 60.5 [49.0, 70.3] |
| DINOv2 | 16.2 [15.4, 17.0] | 32.7 [31.2, 34.2] | 31.4 [30.2, 32.7] | 31.9 [27.2, 33.9] | 48.3 [47.1, 49.4] | 48.9 [42.4, 56.5] |
| VideoMAE V2 | 22.1 [21.1, 23.4] | 35.4 [33.7, 36.8] | 30.4 [29.4, 31.5] | 31.2 [29.8, 32.2] | 48.1 [46.9, 49.3] | 51.4 [45.0, 58.2] |
| V-JEPA 2 | 15.7 [14.7, 16.9] | 29.0 [27.6, 30.3] | 32.1 [31.1, 33.1] | 36.0 [30.0, 38.9] | 44.7 [43.7, 45.7] | 54.9 [46.6, 62.0] |
| PeskaVLP | 17.5 [16.1, 19.1] | 33.1 [31.3, 34.6] | 45.9 [43.6, 48.3] | 43.0 [35.4, 45.9] | 52.4 [51.0, 53.8] | 54.9 [44.0, 63.3] |
| SurgRec-MAE | 13.9 [13.5, 14.4] | 25.6 [24.4, 26.7] | 25.6 [24.8, 26.4] | 23.5 [18.3, 26.5] | 43.5 [42.5, 44.5] | 54.4 [45.7, 62.6] |
| ViSiL | 20.3 [19.0, 21.9] | 40.1 [37.6, 42.2] | 30.5 [29.3, 31.7] | 31.0 [29.5, 32.1] | 50.4 [48.4, 52.3] | 55.2 [46.8, 63.0] |
| C3D + SupCon | 26.2 [24.5, 28.0] | 47.2 [44.6, 49.7] | 42.2 [39.6, 44.9] | 44.4 [37.6, 48.2] | 76.1 [73.9, 78.3] | 46.2 [41.9, 52.0] |
| DINOv2 + SupCon | 47.3 [44.7, 50.2] | 59.8 [57.5, 62.2] | 63.1 [58.8, 67.6] | 64.6 [56.0, 67.5] | 91.3 [89.6, 92.8] | 43.5 [38.4, 50.3] |
| VideoMAE V2 + SupCon | 64.7 [62.3, 67.2] | 68.1 [65.4, 70.6] | 66.7 [63.0, 70.7] | 64.6 [56.1, 68.3] | 90.4 [88.6, 91.8] | 47.2 [41.9, 52.9] |
| V-JEPA 2 + SupCon | 58.0 [55.4, 60.8] | 64.8 [62.2, 67.4] | 65.4 [61.0, 69.4] | 68.1 [62.3, 70.4] | 91.3 [89.2, 93.1] | 51.3 [47.4, 55.9] |
| PeskaVLP + SupCon | 40.9 [37.6, 44.5] | 60.3 [57.7, 62.9] | 64.0 [59.6, 68.6] | 62.5 [53.8, 66.2] | 89.1 [87.5, 90.6] | 50.5 [46.8, 53.8] |
| SurgRec-MAE + SupCon | 39.4 [37.4, 41.3] | 56.3 [53.5, 58.9] | 48.8 [47.0, 50.7] | 61.7 [53.3, 65.2] | 85.4 [82.8, 87.7] | 51.9 [44.7, 59.7] |

Table 18. Classifier oracle (one-hidden-layer MLP probe on frozen features): balanced accuracy (%) on the test split of the training dataset (own) and of the other dataset of the same procedure (other; shared classes only). WF: transition types as classes.

| | C3D | | DINOv2 | | VideoMAE V2 | | V-JEPA 2 | | PeskaVLP | | SurgRec-MAE | |
|---|---|---|---|---|---|---|---|---|---|---|---|---|
| Train dataset | own | other | own | other | own | other | own | other | own | other | own | other |
| Cataract-LMM | 37.2 | 15.3 | 55.6 | 34.2 | 71.7 | 30.0 | 62.4 | 40.3 | 47.3 | 20.1 | 53.2 | 24.2 |
| Cataract-1K | 45.0 | 9.8 | 64.1 | 22.2 | 72.3 | 20.2 | 73.6 | 21.1 | 64.6 | 14.5 | 48.3 | 14.2 |
| MIGS-GT | 57.2 | – | 64.5 | – | 74.8 | – | 70.3 | – | 67.2 | – | 65.7 | – |
| Cholec80 | 64.6 | 42.4 | 80.0 | 61.7 | 84.0 | 63.8 | 81.5 | 61.7 | 83.0 | 57.5 | 71.1 | 54.1 |
| HeiChole | 51.3 | 47.4 | 67.7 | 62.0 | 64.3 | 59.2 | 67.6 | 63.7 | 66.1 | 65.5 | 57.2 | 41.1 |
| GraSP | 41.9 | – | 64.5 | – | 68.0 | – | 74.7 | – | 64.5 | – | 66.0 | – |
| SurgVU | 83.4 | – | 92.0 | – | 91.8 | – | 93.3 | – | 90.7 | – | 89.5 | – |
| REAL-Colon | 54.0 | – | 45.6 | – | 56.5 | – | 54.3 | – | 55.8 | – | 59.0 | – |
| Cataract-LMM (WF) | 8.8 | 11.4 | 12.6 | 16.3 | 19.0 | 13.3 | 13.5 | 11.3 | 11.5 | 9.0 | 15.2 | 14.4 |
| Cataract-1K (WF) | 39.7 | 9.2 | 45.2 | 16.2 | 61.8 | 23.5 | 66.8 | 23.3 | 55.6 | 18.7 | 32.4 | 13.9 |
| MIGS-GT (WF) | 25.9 | – | 34.4 | – | 32.6 | – | 33.6 | – | 37.1 | – | 36.4 | – |

Table 19. Classifier oracle (linear probe on frozen features): balanced accuracy (%) on the test split of the training dataset (own) and of the other dataset of the same procedure (other; shared classes only). WF: transition types as classes.

| | C3D | | DINOv2 | | VideoMAE V2 | | V-JEPA 2 | | PeskaVLP | | SurgRec-MAE | |
|---|---|---|---|---|---|---|---|---|---|---|---|---|
| Train dataset | own | other | own | other | own | other | own | other | own | other | own | other |
| Cataract-LMM | 36.8 | 18.4 | 57.3 | 36.8 | 72.7 | 28.7 | 64.2 | 44.1 | 53.3 | 24.3 | 51.7 | 14.5 |
| Cataract-1K | 41.2 | 11.8 | 68.8 | 23.9 | 81.0 | 19.8 | 72.6 | 16.4 | 63.6 | 16.5 | 52.8 | 16.3 |
| MIGS-GT | 55.6 | – | 64.8 | – | 74.3 | – | 71.3 | – | 68.6 | – | 65.0 | – |
| Cholec80 | 66.0 | 39.1 | 80.4 | 59.1 | 83.2 | 67.5 | 81.4 | 56.4 | 81.0 | 51.4 | 71.1 | 54.5 |
| HeiChole | 48.4 | 44.5 | 64.1 | 55.8 | 67.9 | 58.4 | 64.6 | 65.6 | 61.8 | 64.1 | 57.2 | 51.7 |
| GraSP | 44.3 | – | 66.7 | – | 69.9 | – | 75.8 | – | 68.9 | – | 67.8 | – |
| SurgVU | 82.8 | – | 91.6 | – | 92.0 | – | 93.6 | – | 90.1 | – | 89.6 | – |
| REAL-Colon | 46.3 | – | 51.3 | – | 58.9 | – | 46.9 | – | 53.5 | – | 57.7 | – |
| Cataract-LMM (WF) | 7.2 | 6.3 | 12.7 | 19.4 | 18.9 | 18.4 | 17.5 | 14.8 | 13.5 | 19.1 | 11.9 | 0.0 |
| Cataract-1K (WF) | 39.2 | 5.5 | 52.9 | 18.4 | 60.9 | 17.5 | 69.6 | 17.6 | 56.7 | 17.7 | 40.5 | 23.7 |
| MIGS-GT (WF) | 22.2 | – | 38.0 | – | 33.7 | – | 27.1 | – | 39.8 | – | 40.0 | – |

Table 20. Ranking with a trained classifier vs. a trained retrieval head (SCENE short, within, mAP % averaged over six procedures). feat: cosine of the hidden layer (MLP) or of the logits (linear probe); label: same predicted label, ties broken randomly; prob: cosine of predicted class probabilities.

| | Linear probe | | | MLP probe | | | |
|---|---|---|---|---|---|---|---|
| Encoder | feat | label | prob | feat | label | prob | SupCon |
| C3D | 41.2 | 38.2 | 44.8 | 38.0 | 39.0 | 46.9 | 47.0 |
| DINOv2 | 57.0 | 52.3 | 62.6 | 50.9 | 51.4 | 61.8 | 61.6 |
| VideoMAE V2 | 58.4 | 57.0 | 66.9 | 51.9 | 56.3 | 66.8 | 67.0 |
| V-JEPA 2 | 55.8 | 54.2 | 65.3 | 52.2 | 54.4 | 65.6 | 66.5 |
| PeskaVLP | 52.1 | 51.2 | 60.6 | 50.7 | 50.9 | 60.2 | 61.2 |
| SurgRec-MAE | 49.5 | 46.6 | 56.9 | 43.9 | 45.7 | 57.6 | 57.2 |

Table 21. SupCon head: validation mAP (%) at the selected step, per dataset (WF: transitions). –: validation split too small to select on (REAL-Colon, 18 clips; the head is then trained for the minimum number of steps).

| Encoder | Cataract-1K | Cataract-1K WF | Cataract-LMM | Cataract-LMM WF | Cholec80 | GraSP | HeiChole | MIGS-GT | MIGS-GT WF | REAL-Colon | SurgVU |
|---|---|---|---|---|---|---|---|---|---|---|---|
| V-JEPA 2 | 72.0 | 69.3 | 56.5 | 34.4 | 83.0 | 67.9 | 74.7 | 73.7 | 42.7 | – | 95.5 |
| PeskaVLP | 67.0 | 46.8 | 38.1 | 19.4 | 80.8 | 58.6 | 83.6 | 65.9 | 33.5 | – | 92.5 |
| SurgRec-MAE | 47.5 | 37.4 | 39.7 | 20.8 | 62.8 | 49.3 | 58.2 | 60.0 | 41.1 | – | 91.1 |

## 7. Qualitative examples

Figures 2–22 illustrate the relevance definition of SurgCBVR on the test split. Each row is one retrieval example: a query clip (blue), three *positives* that share its label but come from other cases (green), and three *negatives* with other labels, also from other cases (red). Clips from the query's own case never appear, as they are removed from every gallery during evaluation. The examples are drawn at random with a fixed seed and show the ground-truth relevance, not the ranking of any method. SCENE tiles show the central frame of the 5 s clip; WORKFLOW tiles show three frames at 5%, 50% and 95% of the clip, *i.e*. the end of the first phase, the change and the start of the second phase. Below each tile, the first line gives the label and the second the case, the length of the source window and the resulting playback speed (*e.g*. "32s src (6.4×)": 32 s of surgery rendered into 5 s). Matching clips for every figure are listed in the released `picks_*.csv` files, and corresponding videos are provided as supplementary files. Figures 2–5 give an overview across datasets; Figures 6–13 show SCENE examples per dataset, Figures 14–17 cross-dataset SCENE examples, and Figures 18–22 WORKFLOW examples within and across datasets.

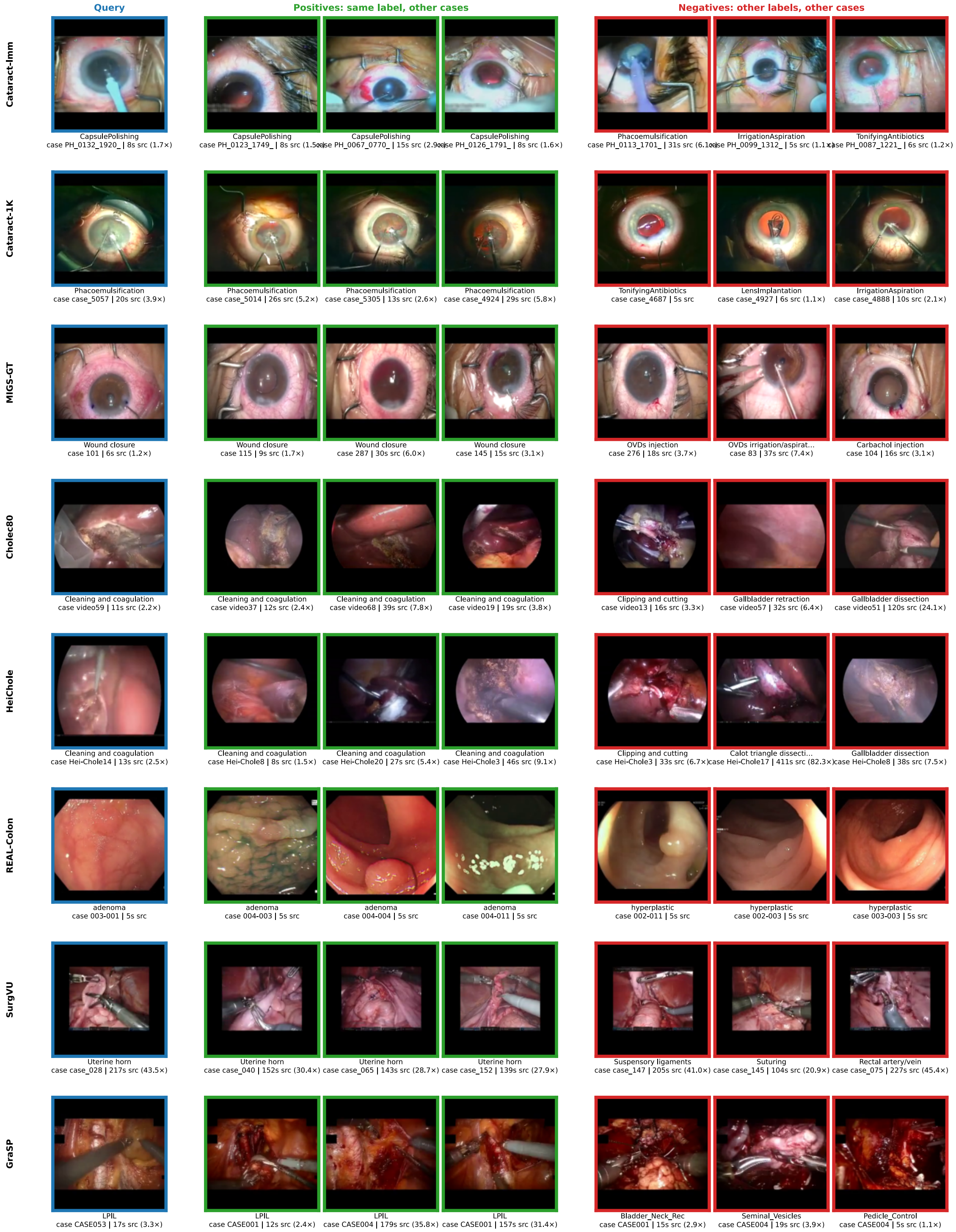

Figure 2. SCENE retrieval examples for all eight datasets (one row each), phase-level clips (5 s rendered from $D/5$ of the source segment); colonoscopy has no phase-level clips and uses short clips. Positives share the query's phase, task or lesion class; negatives are other labels of the same procedure. Within each procedure, positives and negatives often look alike (*e.g.* cleaning and coagulation vs. gallbladder dissection on the same liver bed in Cholec80 and HeiChole, or adenoma vs. hyperplastic polyps in REAL-Colon), which is what makes within-procedure retrieval difficult.

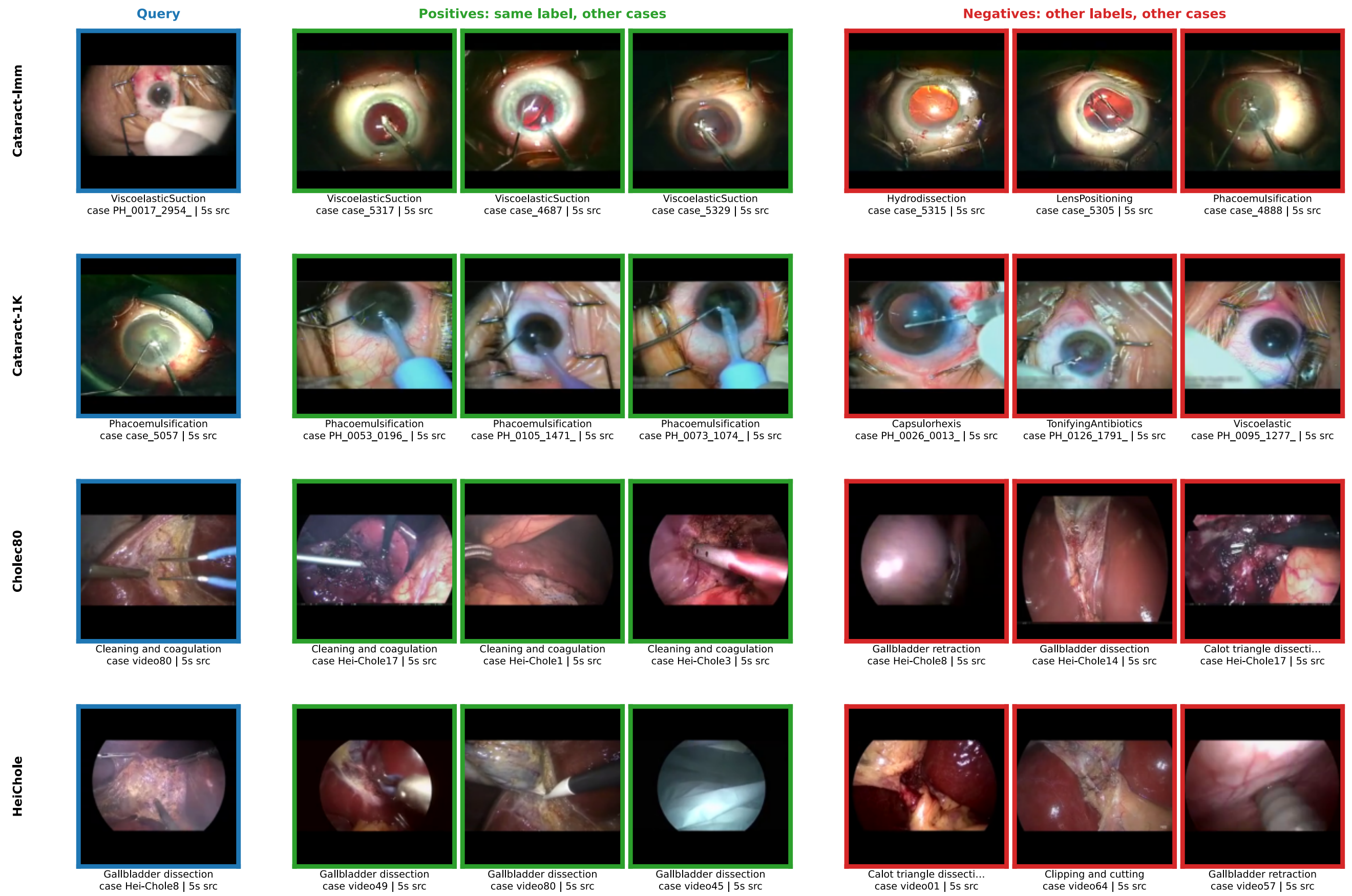

Figure 3. Cross-dataset SCENE retrieval, short clips: queries from one dataset, positives and negatives from the other dataset of the same procedure, restricted to the labels both datasets share (Cataract-LMM ↔ Cataract-1K, Cholec80 ↔ HeiChole). Differences in microscope or laparoscope, illumination, image framing and patient population are visible between query and gallery, while the phase is the same.

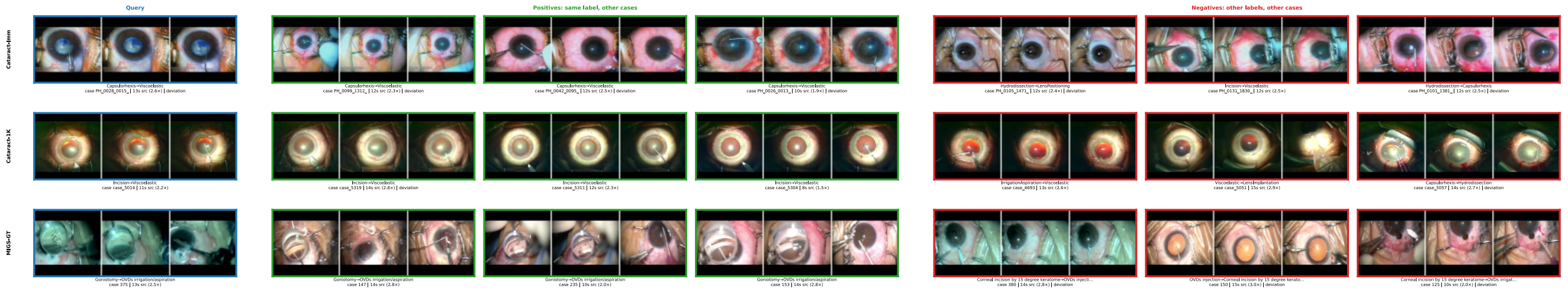

Figure 4. WORKFLOW retrieval examples for the three ophthalmic datasets (one row each). Each tile shows the start, middle and end of a transition clip; positives have the same transition type $A \rightarrow B$, negatives other transition types, which frequently share one of the two phases (*e.g*. Incision→Viscoelastic vs. IrrigationAspiration→Viscoelastic).

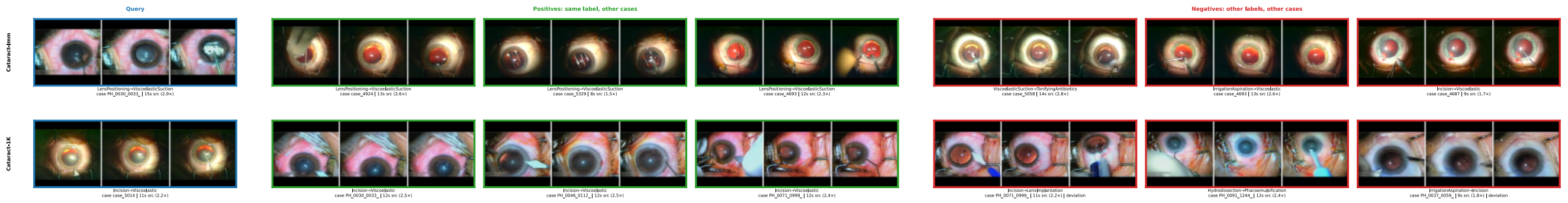

Figure 5. Cross-dataset WORKFLOW retrieval for phacoemulsification: a Cataract-LMM query against Cataract-1K transitions (top) and vice versa (bottom), restricted to transition types present in both datasets.

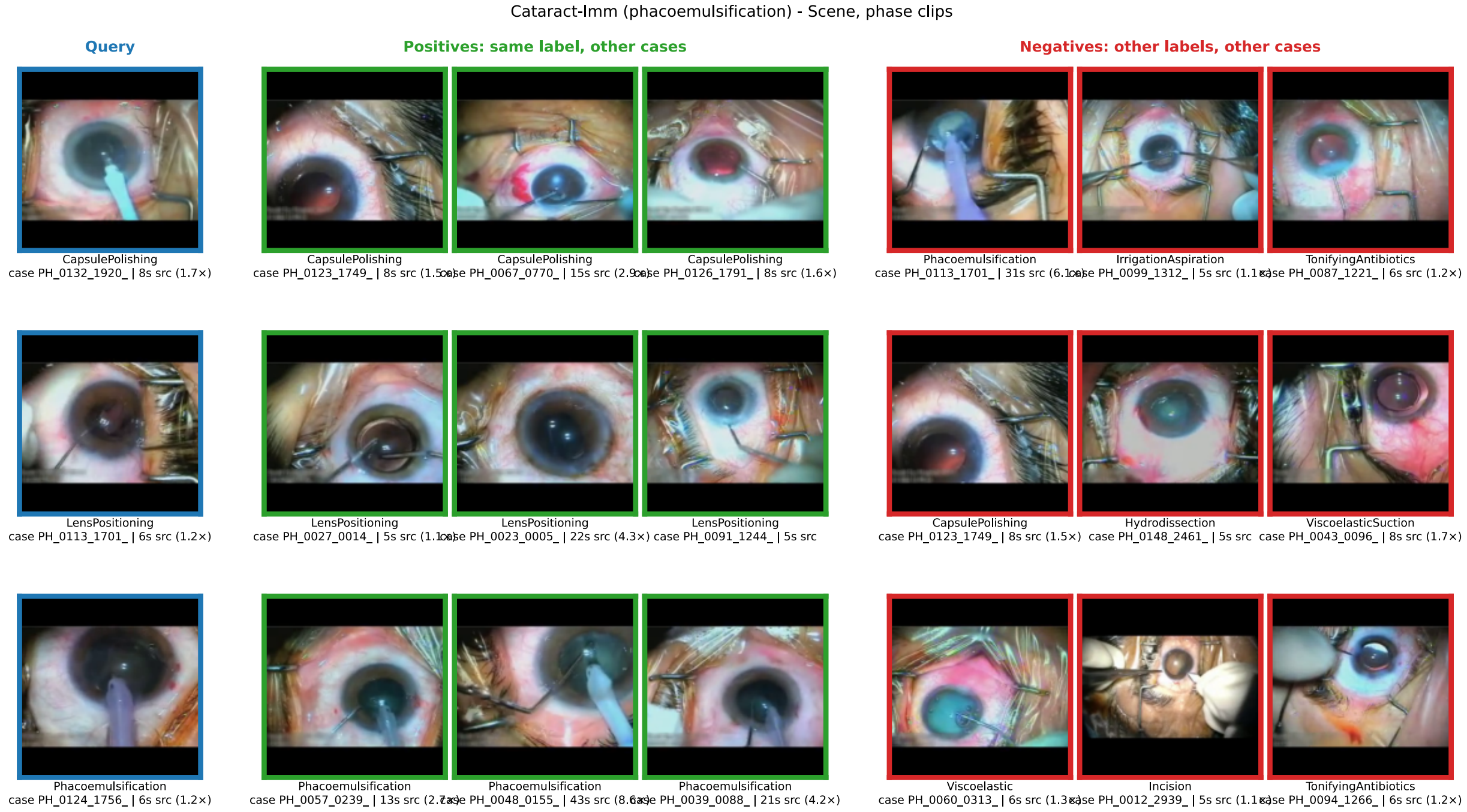

Figure 6. Cataract-LMM, SCENE, phase-level clips. Queries: capsule polishing, lens positioning and phacoemulsification.

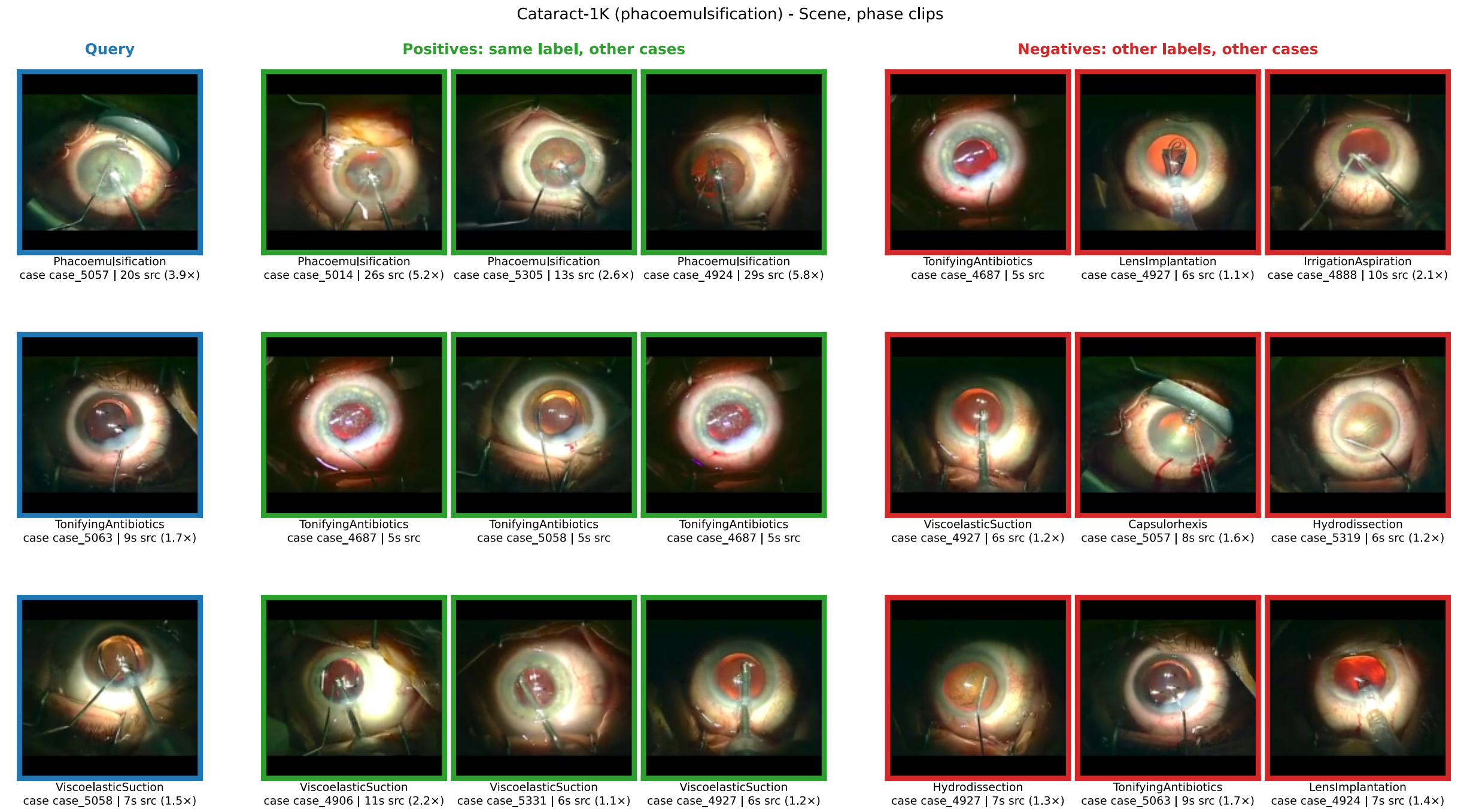

Figure 7. Cataract-1K, SCENE, phase-level clips. Queries: phacoemulsification, tonifying/antibiotics and viscoelastic suction.

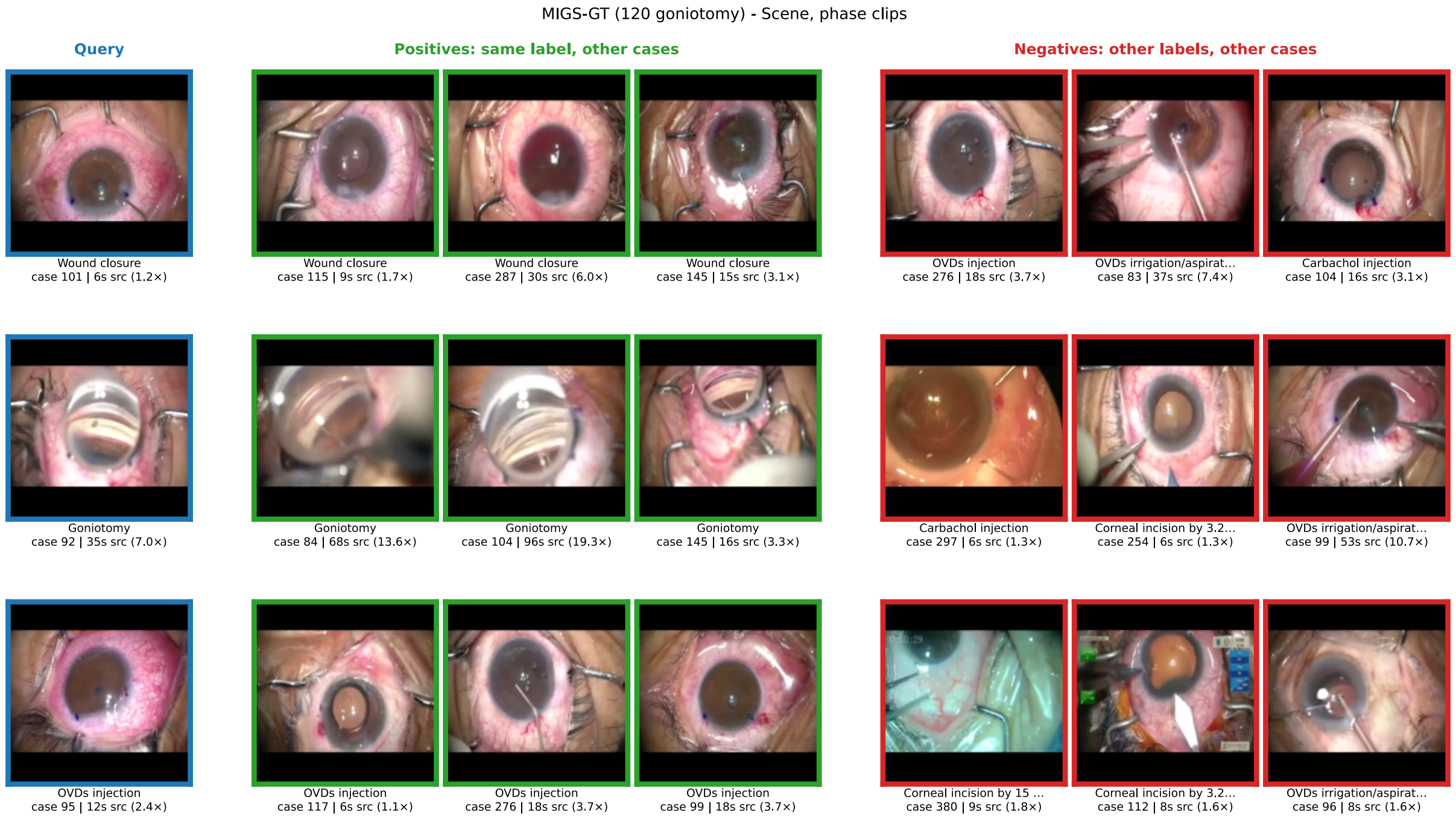

Figure 8. MIGS-GT (goniotomy), SCENE, phase-level clips. Queries: wound closure, goniotomy and OVD injection. Several phases share the same view of the eye and differ mainly in the instrument.

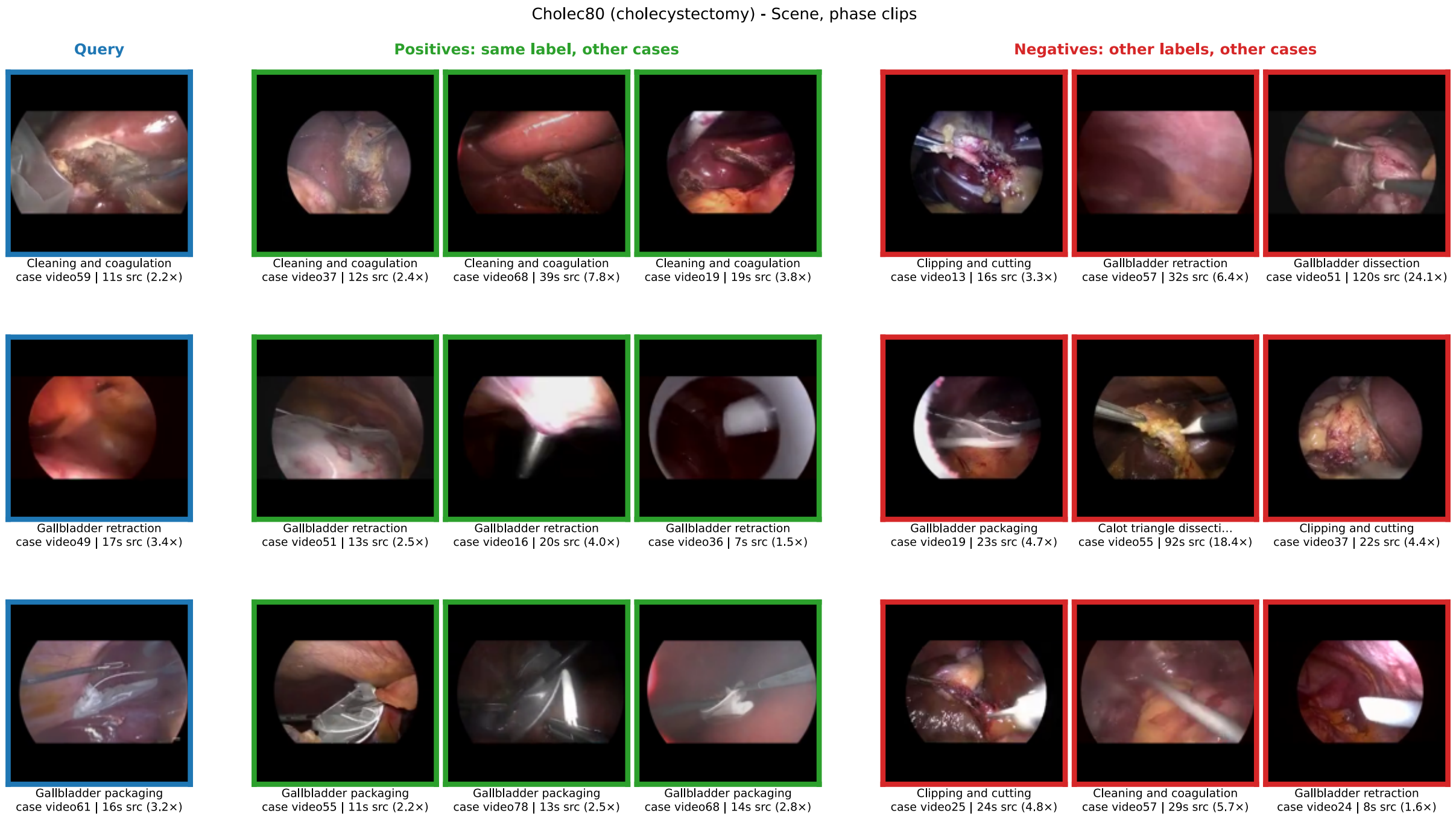

Figure 9. Cholec80, SCENE, phase-level clips. Queries: cleaning and coagulation, gallbladder retraction (extraction of the specimen bag) and gallbladder packaging. Phase-level clips cover longer source windows and are played faster (up to 24× here).

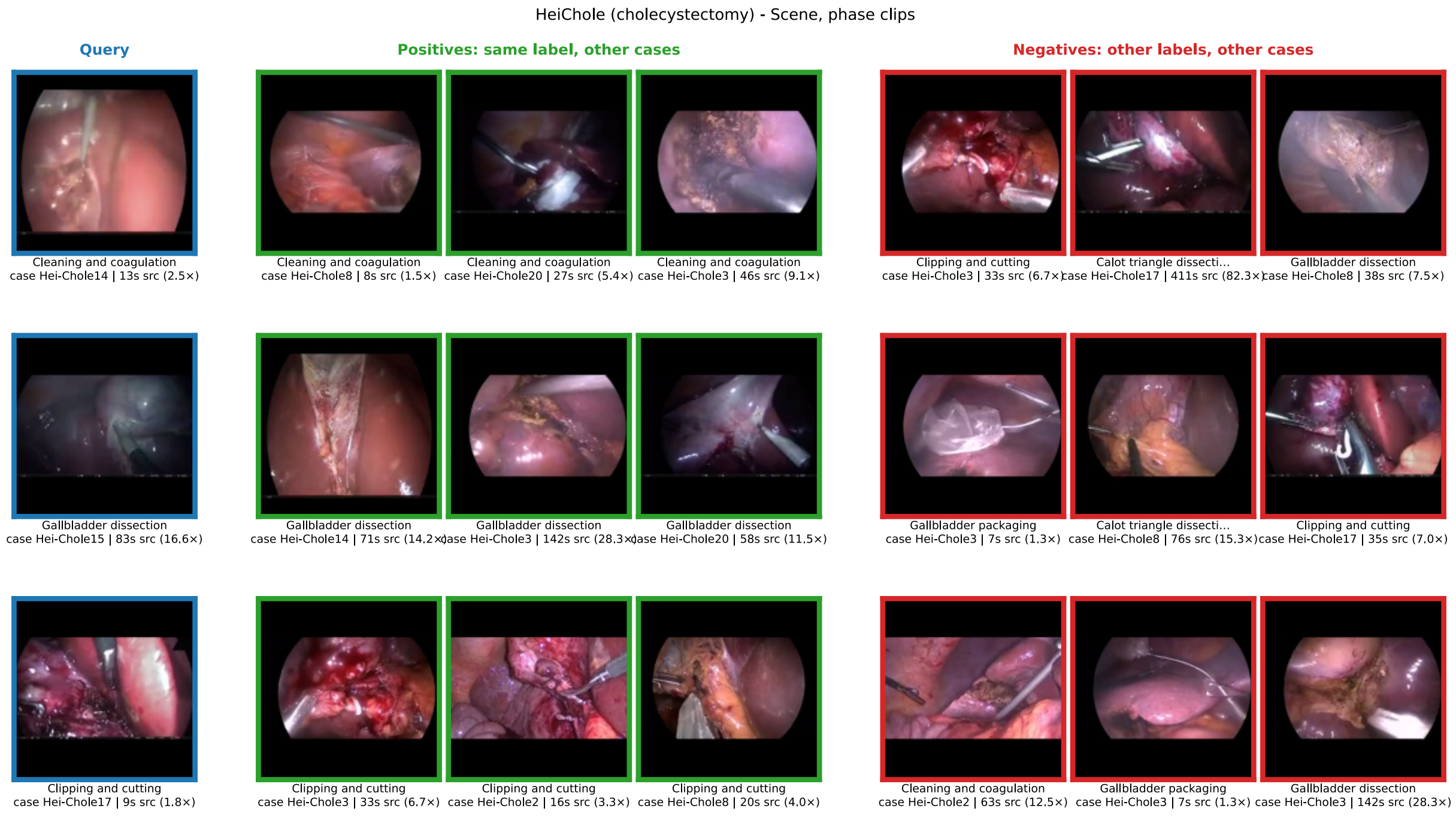

Figure 10. HeiChole, SCENE, phase-level clips. Queries: cleaning and coagulation, gallbladder dissection and clipping and cutting.

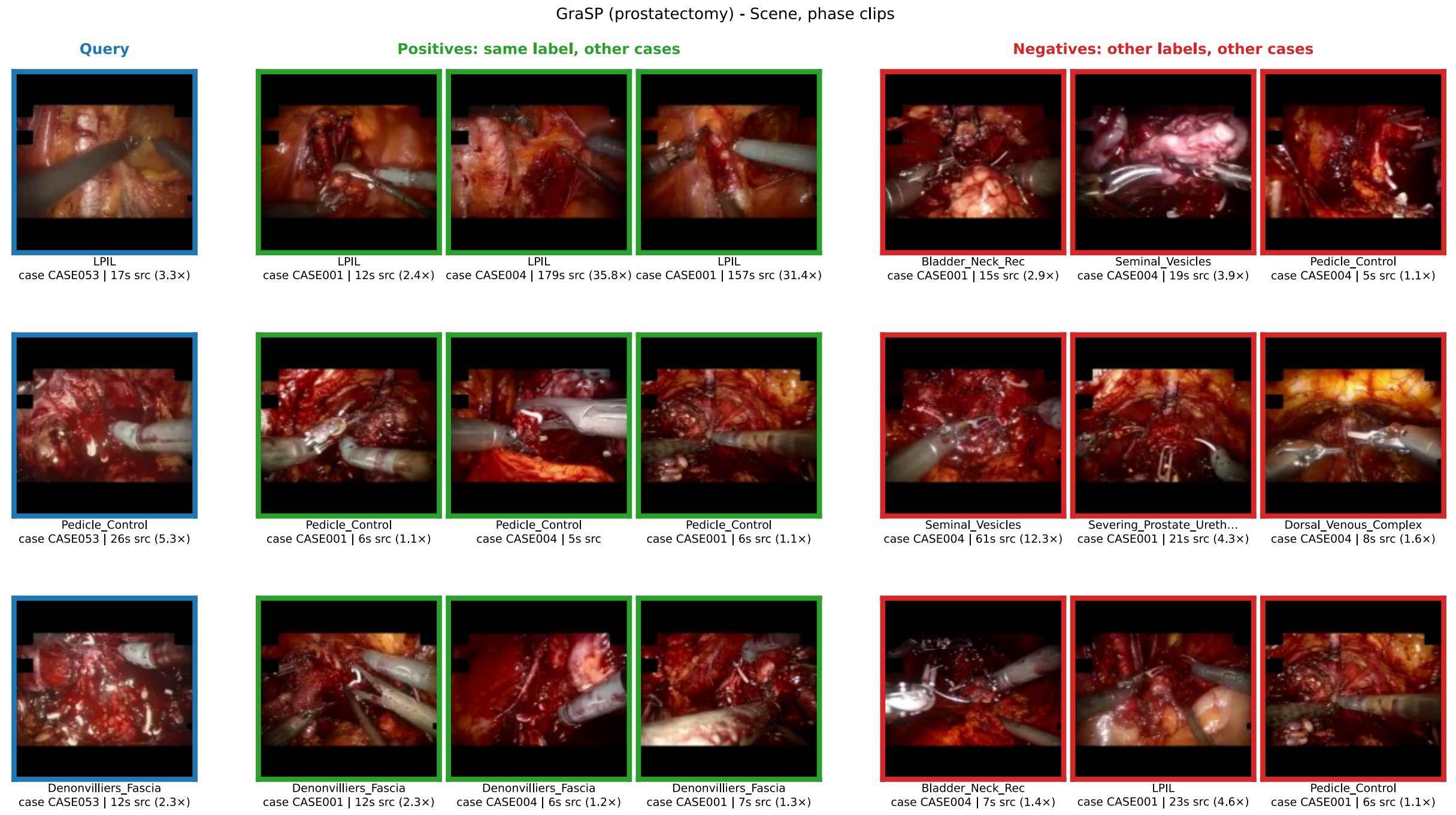

Figure 11. GraSP (robot-assisted radical prostatectomy), SCENE, phase-level clips. Queries: left pelvic isolated lymphadenectomy (LPIL), vascular pedicle control and Denonvilliers' fascia dissection. The test split contains only three patients, so positives can repeat a case.

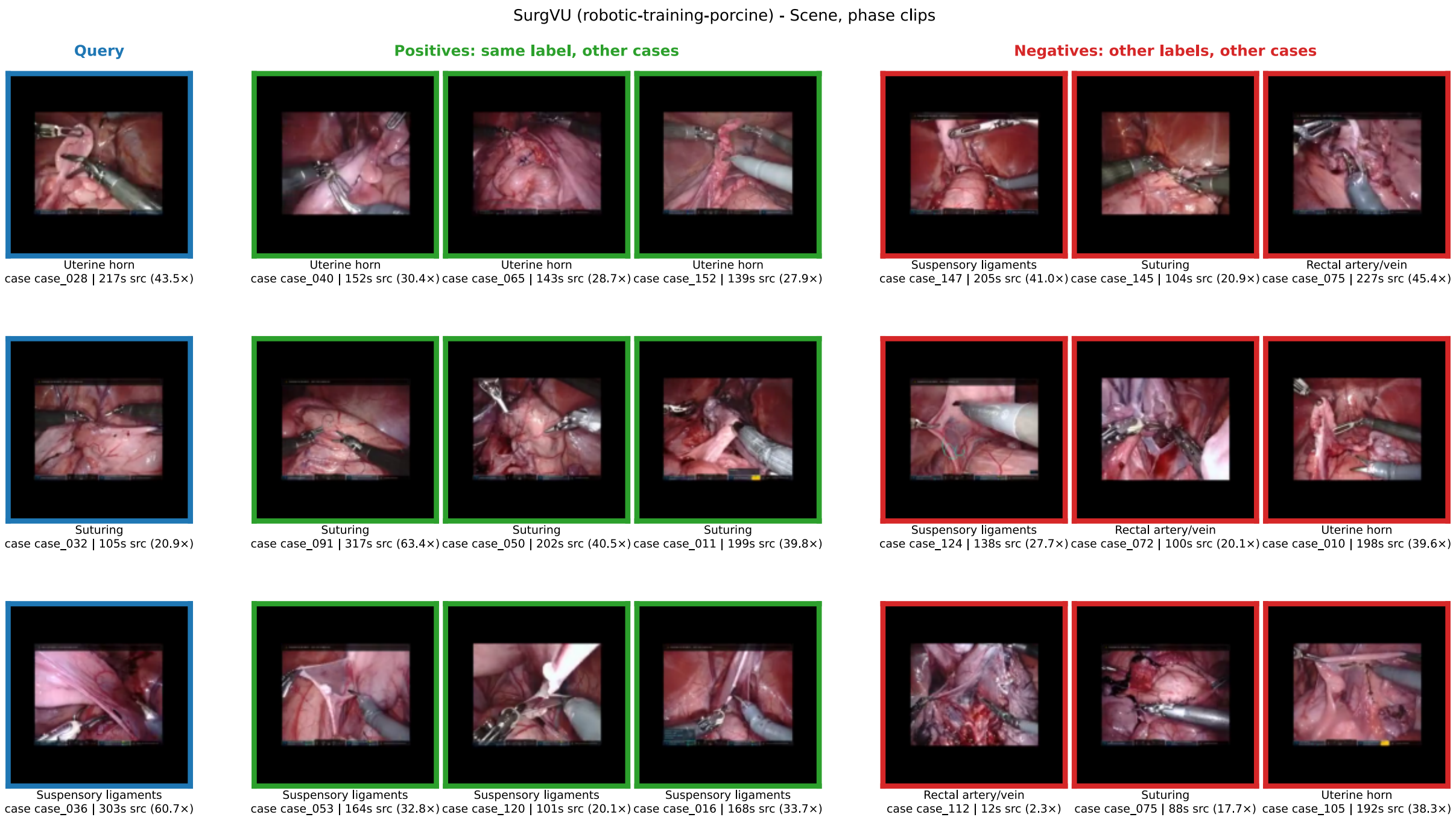

Figure 12. SurgVU (robotic training on porcine models), SCENE, phase-level clips. Queries: uterine horn, suturing and suspensory ligaments. Long task segments give playback speeds above 40× for phase-level clips.

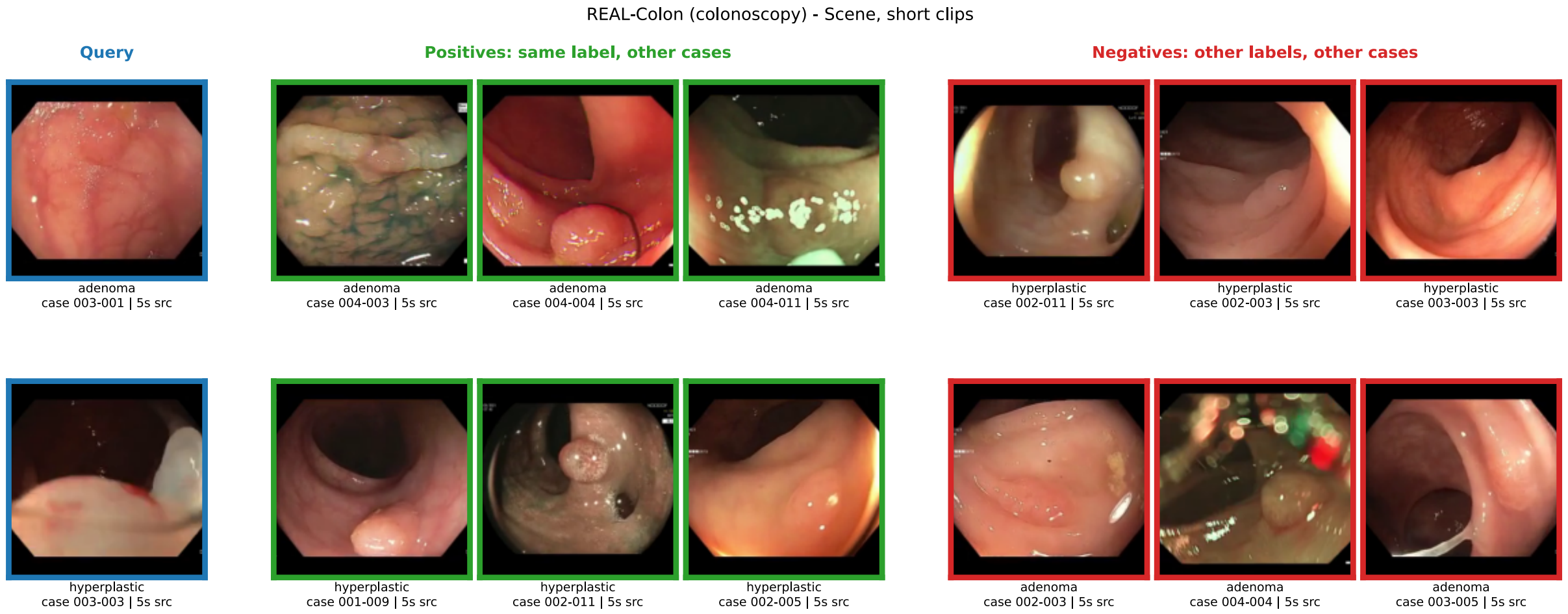

Figure 13. REAL-Colon, SCENE, short clips (colonoscopy has no phase-level clips). Queries: adenoma and hyperplastic polyp; with two classes, every negative is the other class. Histology is hard to judge from a single white-light view, and lesions can be small in the frame.

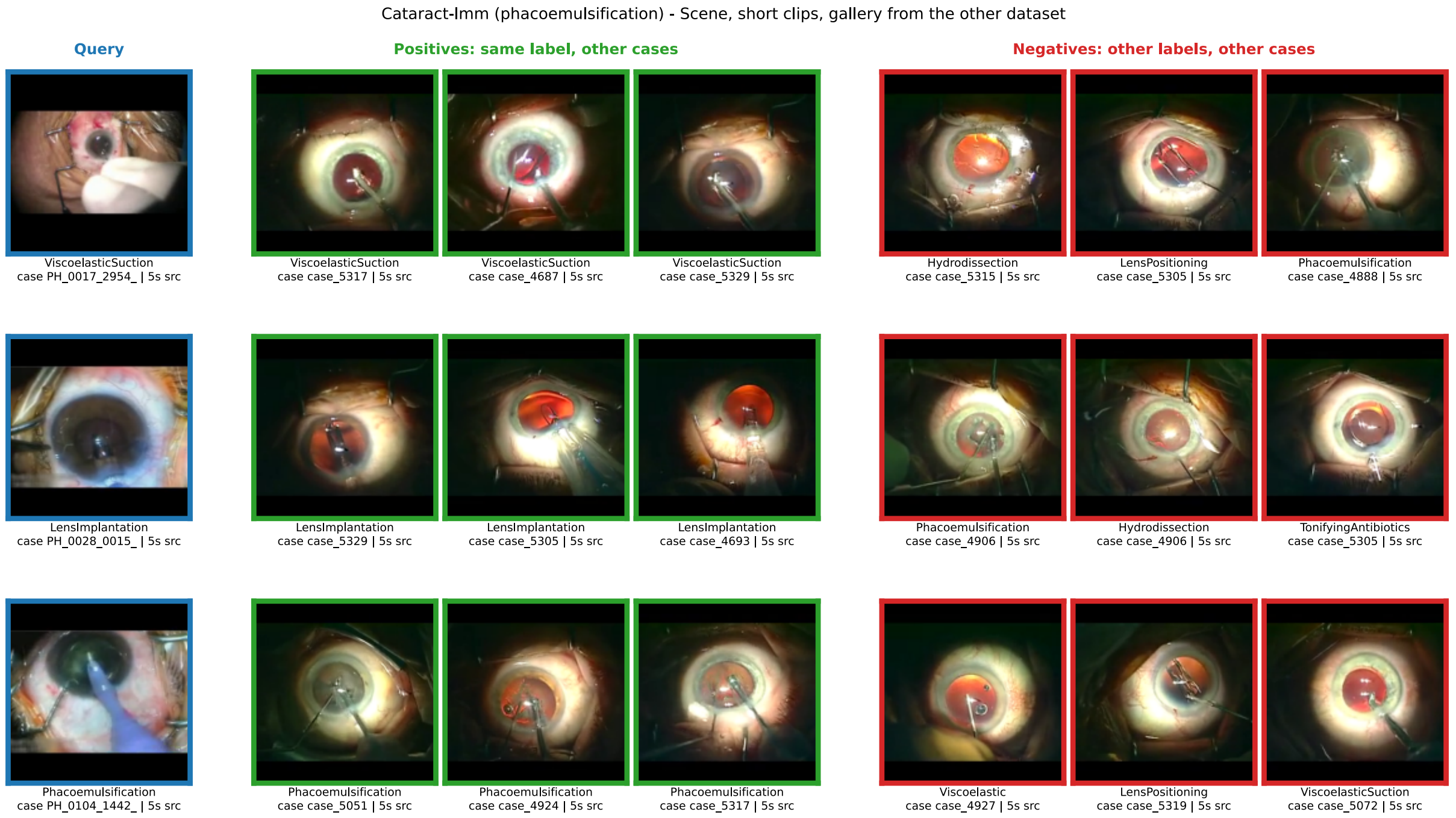

Figure 14. Cross-dataset SCENE retrieval, short clips: Cataract-LMM queries (viscoelastic suction, lens implantation, phacoemulsification) against Cataract-1K positives and negatives.

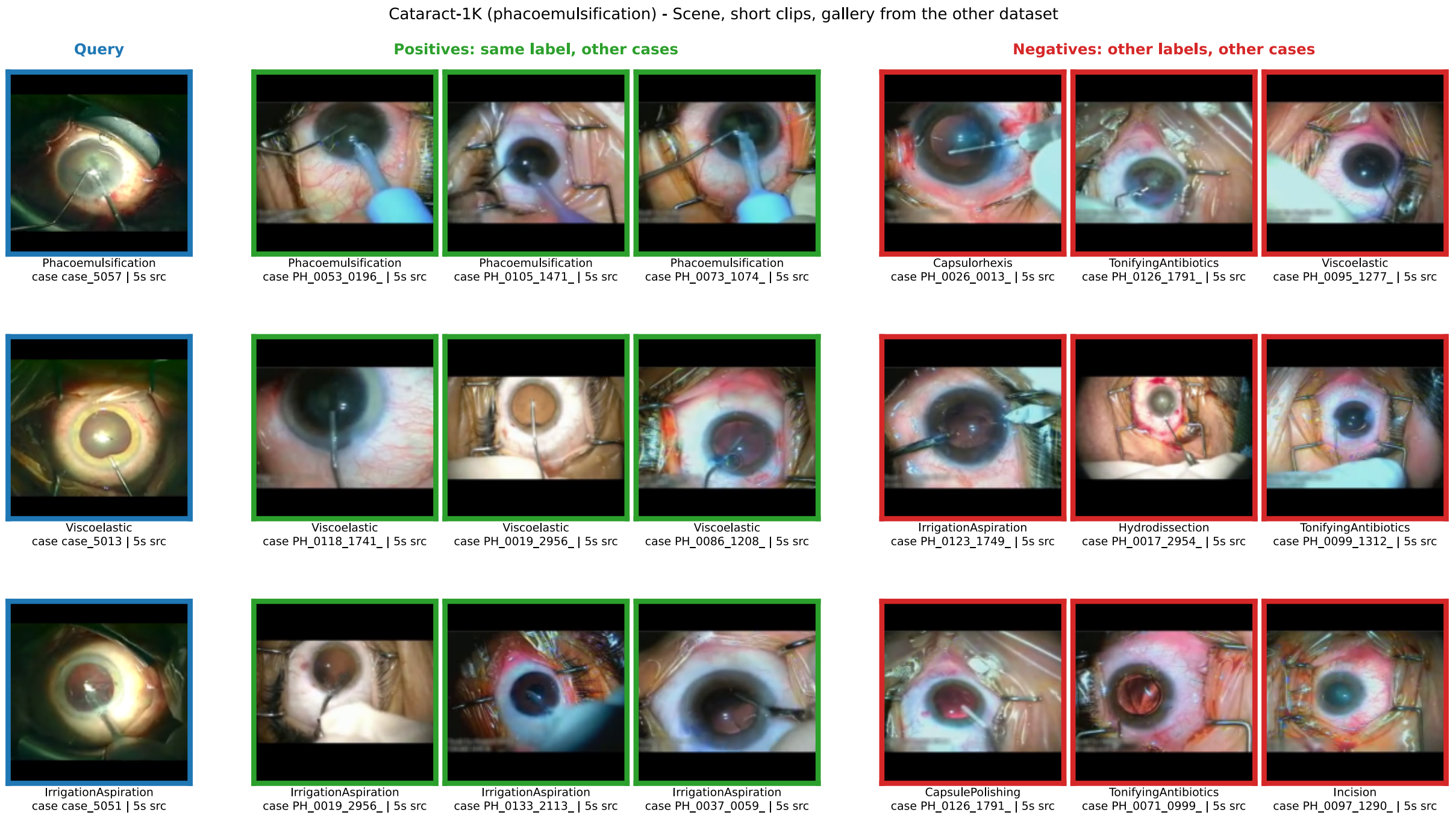

Figure 15. Cross-dataset SCENE retrieval, short clips: Cataract-1K queries (phacoemulsification, viscoelastic, irrigation/aspiration) against Cataract-LMM positives and negatives.

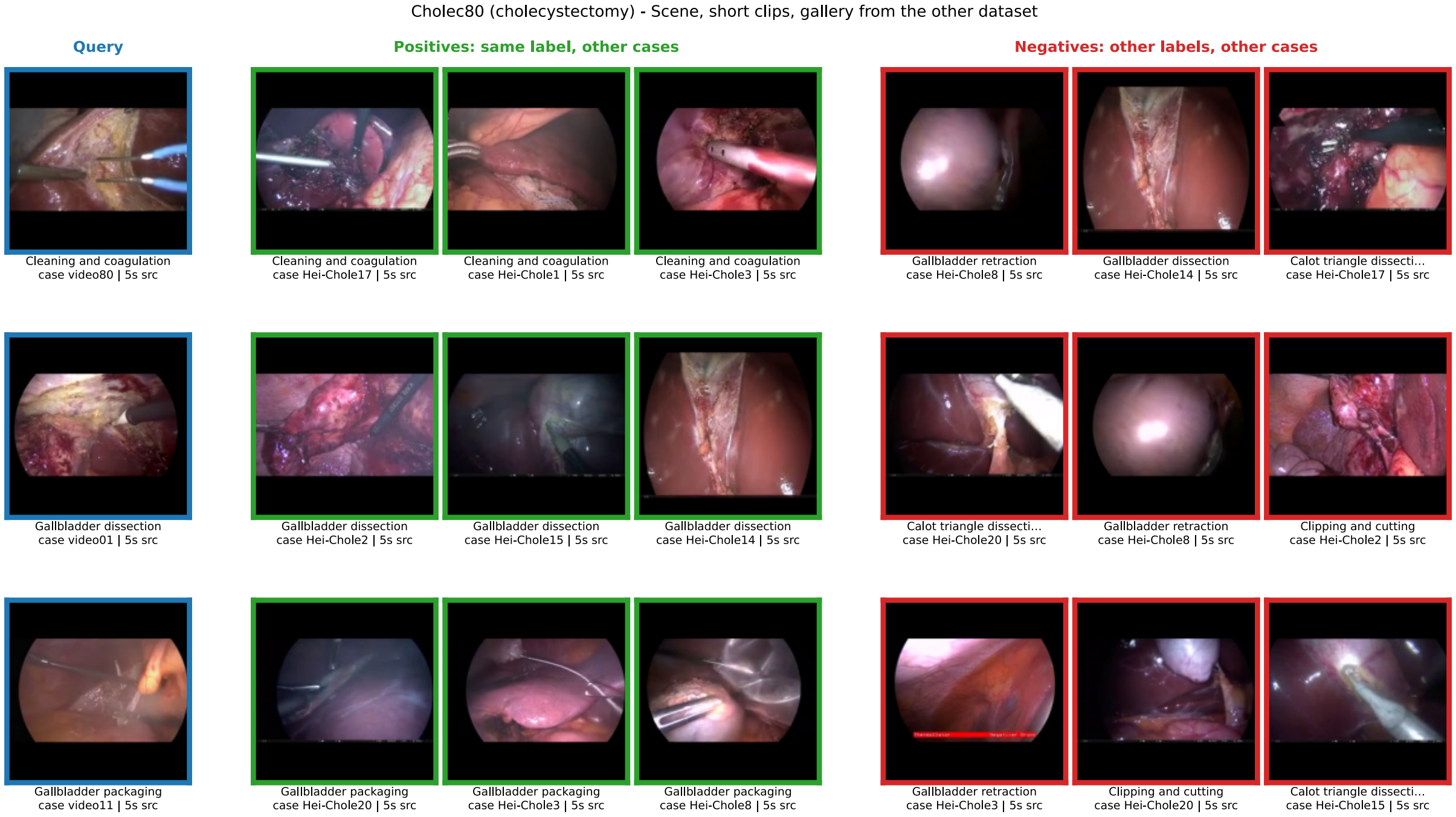

Figure 16. Cross-dataset SCENE retrieval, short clips: Cholec80 queries (cleaning and coagulation, gallbladder dissection, gallbladder packaging) against HeiChole positives and negatives.

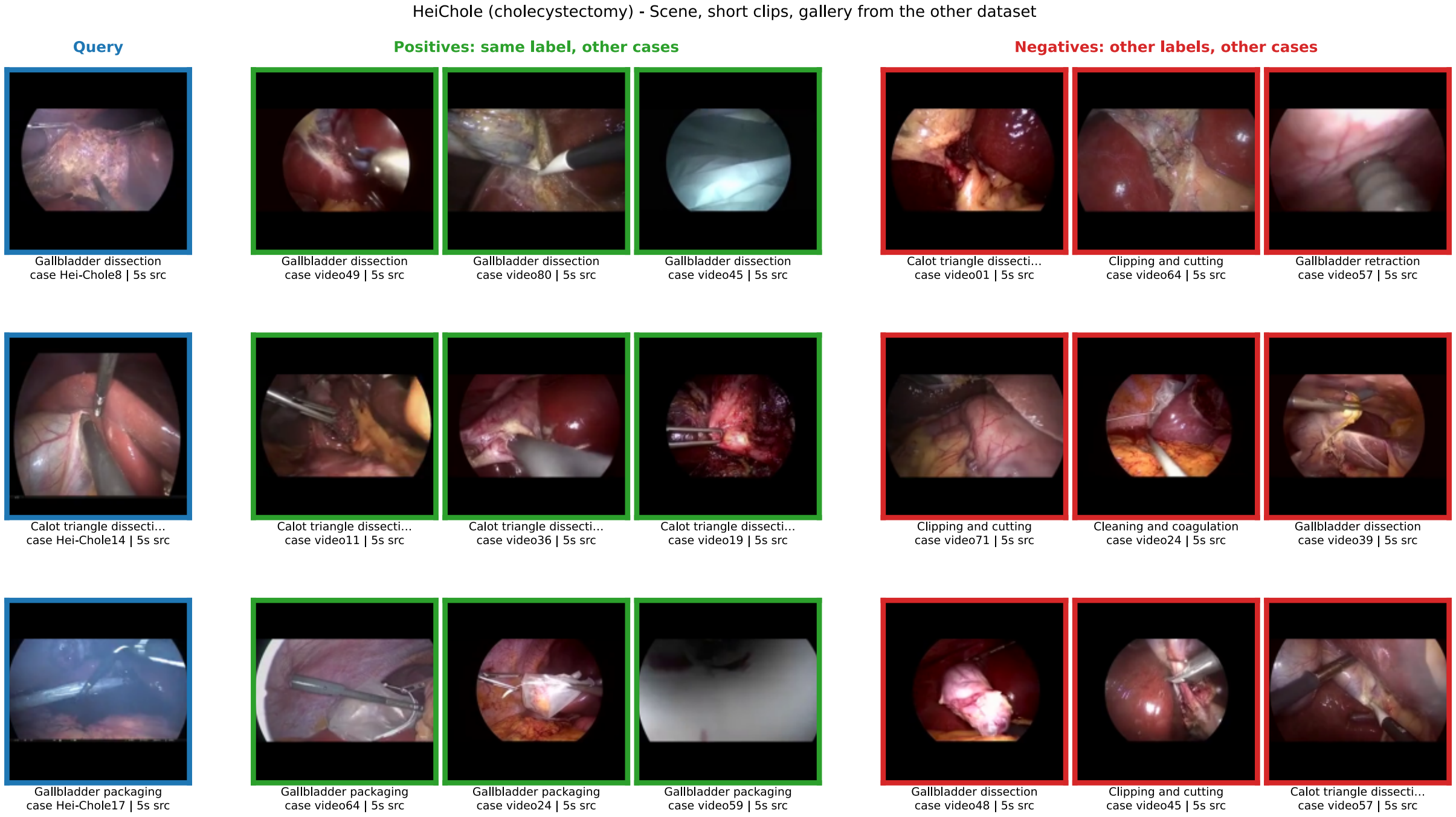

Figure 17. Cross-dataset SCENE retrieval, short clips: HeiChole queries (gallbladder dissection, Calot triangle dissection, gallbladder packaging) against Cholec80 positives and negatives.

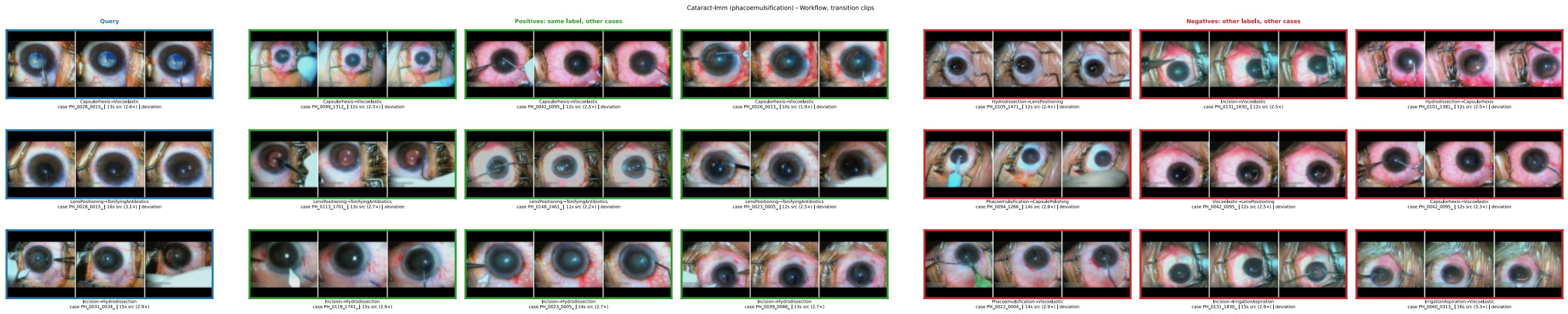

Figure 18. Cataract-LMM, WORKFLOW. Queries: Capsulorhexis→Viscoelastic, LensPositioning→TonifyingAntibiotics and Incision→Hydrodissection. Each tile shows the start, middle and end of the transition window.

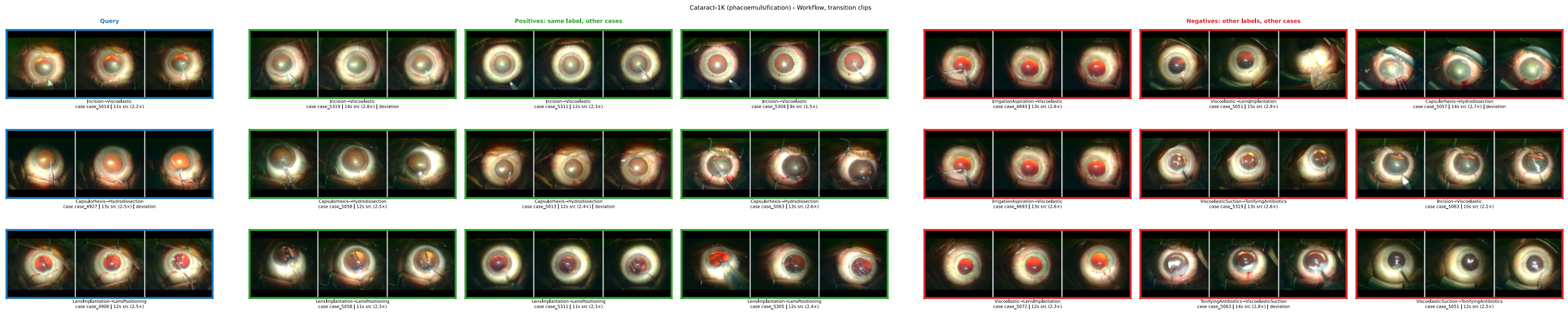

Figure 19. Cataract-1K, WORKFLOW. Queries: Incision→Viscoelastic, Capsulorhexis→Hydrodissection and LensImplantation→LensPositioning.

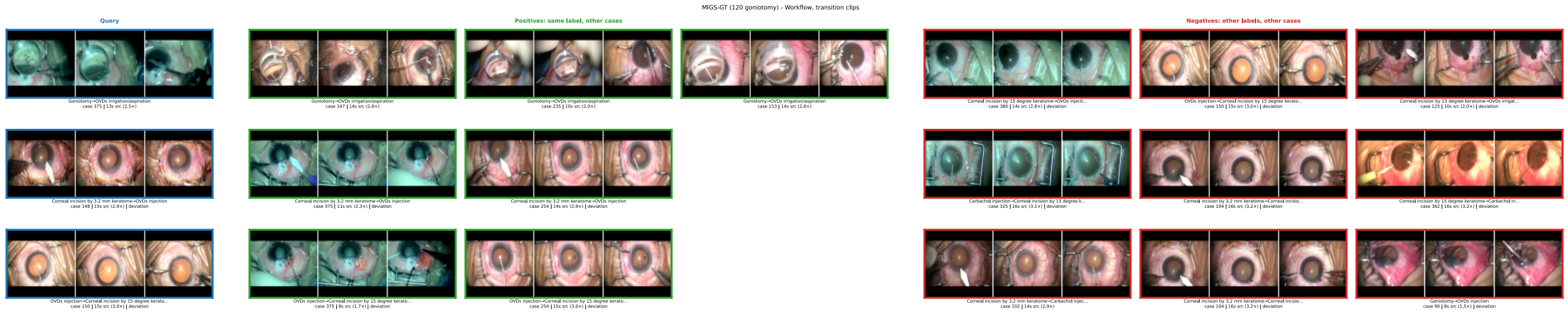

Figure 20. MIGS-GT, WORKFLOW. Queries: goniotomy→OVD irrigation/aspiration, corneal incision (3.2 mm keratome)→OVD injection and OVD injection→corneal incision (15° keratome).

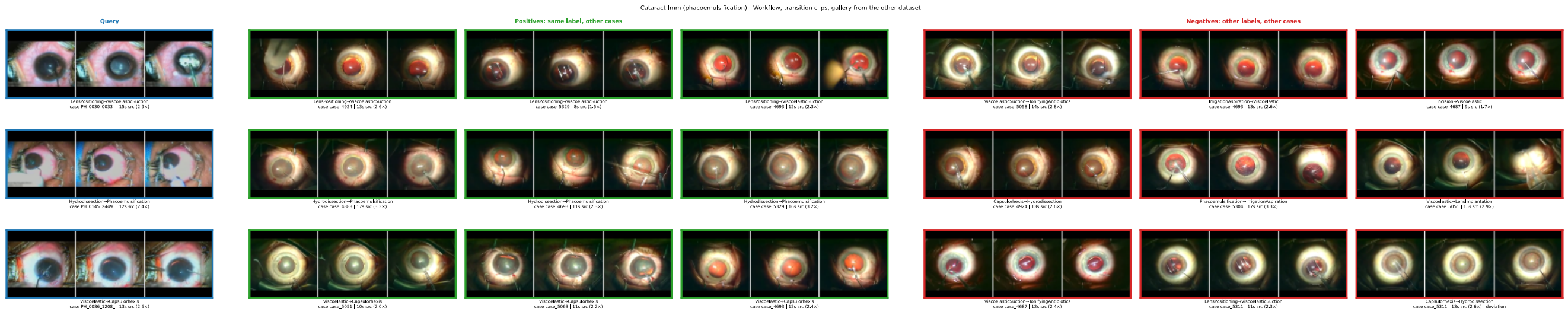

Figure 21. Cross-dataset WORKFLOW: Cataract-LMM queries (LensPositioning→ViscoelasticSuction, Hydrodissection→Phacoemulsification, Viscoelastic→Capsulorhexis) against Cataract-1K transitions.

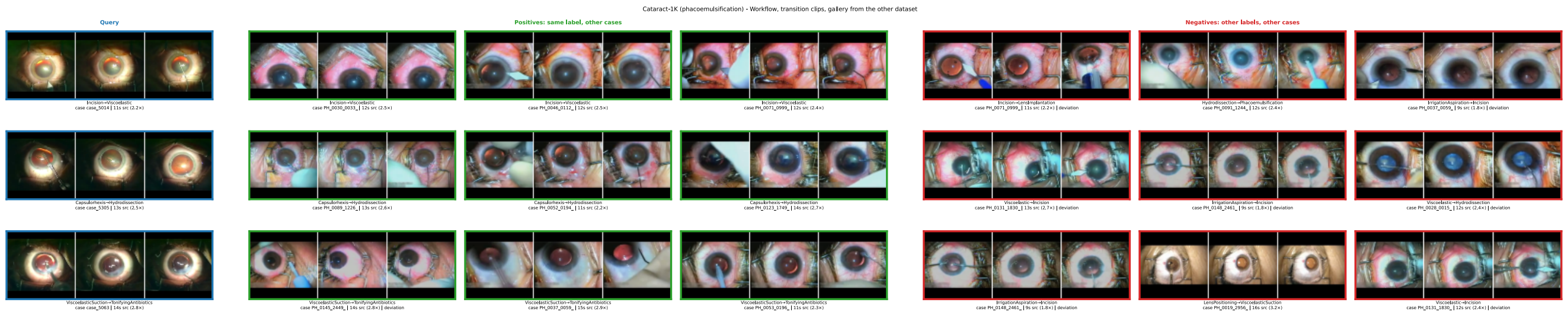

Figure 22. Cross-dataset WORKFLOW: Cataract-1K queries (Incision→Viscoelastic, Capsulorhexis→Hydrodissection, ViscoelasticSuction→TonifyingAntibiotics) against Cataract-LMM transitions.